\documentclass{aa}

\usepackage{graphicx}
\usepackage{longtable}
\usepackage{supertabular}
\usepackage{txfonts}
\usepackage{natbib}
\usepackage{color}
\usepackage{amsmath}
\usepackage{amssymb}	
\usepackage{bm}
\usepackage{float}
\usepackage{dblfloatfix}
\usepackage{textgreek}
\usepackage{array}
\usepackage{textcomp}
\usepackage{upgreek}
\usepackage{cellspace}
\usepackage{caption}
\usepackage{subcaption}
\usepackage{adjustbox}
\usepackage{ulem}

\usepackage{orcidlink}

\bibpunct{(}{)}{;}{a}{}{,} 

\begin{document}

   \title{Characterising the orbit and circumstellar environment of the high-mass binary \object{MWC\,166\,A}}

   \author{Sebastian A. Zarrilli
          \inst{1}\,\orcidlink{0000-0003-4212-8873}
          \and
          Stefan Kraus
          \inst{1}\,\orcidlink{0000-0001-6017-8773}
          \and  
          Alexander Kreplin
          \inst{1}\,\orcidlink{0000-0002-0911-9505}
          \and
          John D.\ Monnier
          \inst{2}\,\orcidlink{0000-0002-3380-3307}
          \and
          Tyler Gardner
          \inst{2}\,\orcidlink{0000-0002-3003-3183}
          \and
          Antoine M\'{e}rand
          \inst{5}\,\orcidlink{0000-0003-2125-0183}
          \and
          Sam Morrell
          \inst{1}\,\orcidlink{0000-0001-6352-5312}
          \and
          Claire L.\ Davies
          \inst{1}\,\orcidlink{0000-0001-9764-2357}
          \and
          Aaron Labdon
          \inst{1,5}\,\orcidlink{0000-0001-8837-7045}
          \and
          Jacob Ennis
          \inst{2}\,\orcidlink{0000-0002-1575-4310}
          \and
          Benjamin Setterholm
          \inst{2}\,\orcidlink{0000-0001-5980-0246}
          \and
          Jean-Baptiste Le Bouquin
          \inst{3}\,\orcidlink{0000-0002-0493-4674}
          \and
          Narsireddy Anugu
          \inst{4}\,\orcidlink{0000-0002-2208-6541}
          \and
          Cyprien Lanthermann
          \inst{4}\,\orcidlink{0000-0001-9745-5834}     
          \and
          Gail Schaefer
          \inst{4}\,\orcidlink{0000-0001-5415-9189}
          \and
          Theo ten Brummelaar
          \inst{4}\,\orcidlink{0000-0002-0114-7915}
          }

   \institute{
   (1) University of Exeter, School of Physics and Astronomy, Astrophysics Group, Stocker Road, Exeter, EX4 4QL, UK\\
   (2) University of Michigan, Department of Astronomy, S University Avenue, Ann Arbor, MI 48109, USA\\
   (3) Institut de Planetologie et d'Astronomie de Grenoble, Grenoble 38058, France\\
   (4) The CHARA Array of Georgia State University, Mount Wilson Observatory, Mount Wilson, CA 91023, USA\\
   (5) European Organisation for Astronomical Research in the Southern Hemisphere (ESO), Karl-Schwarzschild-Str. 2, 85748 Garching bei M\"{u}nchen, Germany
   }

   \date{Received \textit{27 May, 2022}; accepted \textit{30 June, 2022}}

  \abstract
   {
   Stellar evolution models are highly dependent on accurate mass estimates, especially for highly massive stars in the early stages of stellar evolution. The most direct method for obtaining model-independent stellar masses is derivation from the orbit of close binaries.
   }
   {
   Our aim was to derive the first astrometric+radial velocity orbit solution for the single-lined spectroscopic binary star MWC\,166\,A, based on near-infrared interferometry over multiple epochs and \textasciitilde100 archival radial velocity measurements, and to derive fundamental stellar parameters from this orbit. A supplementary aim was to model the circumstellar activity in the system from $K$-band spectral lines.
   }
   {
   The data used include interferometric observations from the VLTI instruments GRAVITY and PIONIER, as well as the MIRC-X instrument at the CHARA Array. We geometrically modelled the dust continuum to derive relative astrometry at 13 epochs, determine the orbital elements, and constrain individual stellar parameters at four different age estimates. We used the continuum models as a base to examine differential phases, visibilities and closure phases over the Br\,$\gamma$ and He\,\textsc{i} emission lines, in order to characterise the nature of the circumstellar emission.
   }
   {
   Our orbit solution suggests a period of $P=367.7\pm0.1$~\,d, approximately twice as long as found with previous radial velocity orbit fits. We derive a semi-major axis of $2.61\pm0.04$\,au at $d=990\pm50$\,pc, an eccentricity of $0.498\pm0.001$ and an orbital inclination of $53.6\pm0.3^{\circ}$. This allowed constraint of the component masses to $M_1=12.2\pm2.2\,M_\odot$ and $M_2=4.9\pm0.5\,M_\odot$.
   }
   {The line-emitting gas was found to be localised around the primary and is spatially resolved on scales of $\sim 11$ stellar radii, where the spatial displacement between the line wings is consistent with a rotating disc.  The large spatial extent and stable rotation axes orientation measured for the Br\,$\gamma$ and He\,\textsc{i} line emission are inconsistent with an origin in magnetospheric accretion or boundary-layer accretion, but indicate a ionised inner gas disk around this Herbig Be star.
   We observe line variability that could be explained either with generic line variability in a Herbig star disc or V/R variations in a decretion disc scenario. We have also constrained the age of the system, with relative flux ratios suggesting an age of $\sim(7\pm2)\times10^5$ yr, consistent with the system being comprised of a main-sequence primary and a secondary still contracting towards the main-sequence stage.
   }


   \maketitle
%

\section{Introduction}
\label{sec:intro}

The masses and ages of young stellar objects (YSOs) are commonly derived from comparison of observed positions on a colour-magnitude diagram to theoretical models \citep[e.g.][]{siess00,baraffe15,mesa_tracks}. Evolutionary tracks of YSOs are very sensitive to mass, and as such need to be calibrated from observed systems with well-constrained masses. The relative paucity of higher-mass Herbig Ae/Be YSOs {($\gtrsim5M_\odot$)} compared to their lower-mass T\,Tauri counterparts means that models of higher-mass stars are less thoroughly calibrated. The high effective temperature of Herbig Be stars ($\gtrsim\,15000$ K), as well as their often-uncertain ages due to mass loss from stellar winds, make it much less straightforward to calculate their masses \citep{massey12}. \citet{stassun14} found that predicted and measured masses can differ by \textasciitilde10\%.

The gold standard for deriving model-independent masses is by taking advantage of the orbital mechanics of binary systems. If both astrometric and radial velocity (RV) data are used, the derived orbital parameters can be combined with reliable distance estimates (if any exist) to extract dynamical masses for the individual objects. This requires observation of a binary system at multiple epochs spread over a substantial fraction of the orbit, so targets with relatively short orbits and small separations are the best candidates. The need for precise astrometry on very small angular scales (\textasciitilde1 mas) has hugely benefitted from the relatively recent development of optical and near-infrared (NIR) interferometry, which has `unlocked' a larger tranche of suitable systems compared to even fifteen years ago.

A typical feature of YSOs is the presence of substantial amounts of circumstellar material left over from stellar formation, taking the form of a disc due to conservation of angular momentum. In single systems, dispersal of the disc occurs from a combination of accretion onto the star, depletion from stellar wind, and condensation into protoplanets, with a typical disc lifetime of 1-3\,Myr \citep{li16}.
However, the picture is more complicated when binary systems are concerned. If the binary is widely separated, each individual star can host its own circumstellar disc, but for many luminous and close Herbig Ae/Be binaries, a single circumbinary disc is the only possible structure for circumstellar material \citep{Pichardo05}. This is due to dynamical interactions between the stars and the disc, which can affect both the accretion properties of the system and the disc's shape and lifetime, although it is unclear under what conditions they will either delay or accelerate disc dispersal \citep{cieza09}.
Dynamical truncation will affect the potential of the disc to form planetary systems by removing or rearranging the material available for planet formation. The known population of circumbinary planets has been substantially increased by Kepler observations \citep[e.g.][]{doyle11}, and numerical simulations suggest that features rare in planets around single stars, such as large eccentricities and planet-star misalignment, are more common in circumbinary systems \citep{chen19}. Further studies on well-characterised young multiple star systems are also essential to study other dynamical mechanisms that might shape the architecture of exoplanetary systems, for instance by moving disc material onto oblique orbits \citep{kraus_gwori}.

The focus of this study is the multiple system MWC\,166 (= HD\,53367, HIP\,34116, V750\,Mon). This is a hierarchical triple system, with a close spectroscopic binary (MWC\,166\,A) orbited by a wide companion (MWC\,166\,B) at a separation of {0.6\arcsec} \citep{fabricius02}. The radial velocity (RV) variations of the spectroscopic binary have first been reported by \citet{Finkenzeller84} and tentative spectroscopic orbit solutions have been presented by \citet{Corporon99} and \citet{Pogodin06}. In this paper, we focus on this inner spectroscopic binary, the components of which have been labelled MWC\,166\,Aa and MWC\,166\,Ab throughout.

MWC\,166 is located in the nearby OB association Canis Major OB1, whose age has been estimated to be $\sim3$\,Myr \citep{Claria74a}. The object also features significant mid-infrared to millimetre excess, which is indicative of a disc around MWC\,166\,A.  The distance to the OB association has been estimated to  $1150\pm140$\,pc \citep{Claria74a}. Subsequent photometric measurements taken from a larger number of sources broadly agree with this value and have more tightly constrained it to $990\pm50$\,pc \citep{Shevchenko99,kaltcheva00}. 

Here, we present near-infrared interferometric observations obtained with the Very Large Telescope Interferometer (VLTI) and the Center for High-Angular Resolution Astronomy \citep[CHARA;][]{tenBrum05} Array, which have allowed us to derive a first astrometric orbital solution of the system. We present our observations in Sect.~\ref{sec:MWCobs}, followed by a discussion of our modelling approach (Sect.~\ref{sec:modelling}). We derive the orbit solution and dynamical mass constraints in Sect.~\ref{sec:results}, while spectral line analysis results are presented in Sect.~\ref{sec:PMOIRED_results}. A discussion on the distribution of circumstellar material -- both in the dust continuum and in prominent $K$-band emission lines -- is presented in Sect.~\ref{sec:discussion}, and our conclusions are summarised in Sect.~\ref{sec:conclusions}.

\section{Observations}
\label{sec:MWCobs}

Near-infrared interferometric observations were taken over a period of 8 years, mainly using the PIONIER \citep{pionier} and GRAVITY \citep{gravity} 4-beam combiners at the VLTI. All VLTI observations employed the 1.8-metre Auxiliary Telescopes. Longer baselines were provided by 4-telescope observations using the CHARA Array instrument MIRC-X \citep{mircx,anugu20}.

The GRAVITY observations were taken in the $K$-band (1.99-2.45\,{\textmu}m) as part of ESO programme 098.C-0910(A). GRAVITY observations include data from the fringe tracker, which operates at a spectral resolution of $\mathcal{R} = \Delta\lambda/\lambda\sim22$, as well as the science combiner either in Medium ($\mathcal{R}\sim500$) or High ($\mathcal{R}\sim4000$) resolution \citep{gravity}. Our observations achieved an angular resolution up to $(\lambda/2B_\mathrm{max})=1.6$\,milliarcseconds (mas) on the longest baselines ($B_\mathrm{max}=130$\,m in length), sufficient to spatially resolve the components of MWC\,166\,A \citep{gravity}. The reduction pipeline used was the GRAVITY data reduction pipeline\footnote{Available at: \url{https://ftp.eso.org/pub/dfs/pipelines/instruments/gravity/gravity-pipeline-manual-1.5.4.pdf}} running in the \texttt{ESOreflex v2.9.1} environment \citep{esoreflex}. Besides the statistical uncertainties computed by the GRAVITY pipeline, we include 5\% and $1^{\circ}$ errors, for the visibility and closure phase respectively, to account for calibration uncertainties.

The {PIONIER} observations covered the $H$-band (1.59-1.75~{\textmu}m) and were obtained as part of multiple ESO programmes: 102.C-0701, 104.C-0737 and 106.21JU. These data were recorded over 6 channels at spectral resolution $\mathcal{R}\sim40$. The reduction pipeline used was \texttt{pndrs v3.52} \citep{pionier}. We also included published data from the large programme 190.C-0963\footnote{Taken from the Optical interferometry DataBase (OiDB), available at: \url{http://oidb.jmmc.fr}.} \citep[e.g.][]{lazareff17,Kluska16}, over 3 spectral channels and with a resolution of $\mathcal{R}\sim15$. 

{MIRC-X} was used in its $H$-band mode as part of the programme 2020B-M7, using 4 of the 6 CHARA Array telescopes. CHARA's much longer maximum baseline of 330\,m allowed us to probe the object geometry at nearly 3-times higher resolution than possible with VLTI. The \texttt{MIRC-X v0.9.5} pipeline\footnote{Available at: \url{https://gitlab.chara.gsu.edu/lebouquj/mircx_pipeline}} \citep[][§ 4]{anugu20} was used to reduce the data.

A full description of the observations is provided in Table~\ref{tab:MWCobs}. Each observation was calibrated by observing suitable calibrator stars with known uniform disc diameters \citep[UDDs, taken from][]{Bourges17}, to account for atmospheric absorption and instrument response. The calibrators were also inspected for signatures of binarity to ensure only single stars were observed. In order to improve the $(u,v)$-coverage, we grouped the individual measurements into epochs. However, this proved to be difficult due to the rapidly-changing orbit of the system. Based on the literature RV orbital period of \textasciitilde183 days \citep{Pogodin06}, a variation of about one degree in position angle (PA) per day is to be expected. Considering that the PA uncertainties in our binary model fits are on the same order of magnitude as this (see Table~\ref{tab:bg_models}), each epoch should include data from at most two consecutive nights, ensuring that the relative positions of the two components do not change significantly during each epoch. The observations for which this consolidation was performed are marked accordingly in Table~\ref{tab:MWCobs}.

\begin{table*}
    \caption{\label{tab:MWCobs} Full list of interferometric observations of MWC\,166. Data from programme 190.C-0963 are lacking calibrator information, due to being taken pre-calibrated from the JMMC OiDB.}
    \centering
    \begin{tabular}{l|c|r|c|c|c}
    \hline
    Date & Programme ID & Array Config & Instrument & $\Delta\lambda/\lambda$ & Calibrator(s) used\\
    \hline
    2013-01-27 & 190.C-0963(A) & K0-A1-G1-J3 & PIONIER &  15 & - \\
    2013-02-20 & 190.C-0963(B) & D0-G1-H0-I1 & PIONIER &  15 & - \\
    \hline
    2017-03-14 & 098.C-0910(A) & A0-G1-J2-J3 & GRAVITY &  22 \& 4000 & HD\,49647  \\
    2017-04-27\,\tablefootmark{a} & 098.C-0910(A) & A0-G1-J2-J3 & GRAVITY & 22 \& 4000 & HD\,57087  \\
    2017-04-28\,\tablefootmark{a} & 098.C-0910(A) & A0-G1-J2-J3 & GRAVITY & 22 \& 500 & HD\,49647 \\
    2018-01-11 & 098.C-0910(A) & A0-G1-J2-J3 & GRAVITY &  22 \& 4000 & HD\,49647 \\
    2018-02-06 & 098.C-0910(A) & A0-G1-J2-J3 & GRAVITY &  22 \& 4000 & HD\,38117 \\
    \multicolumn{1}{c|}{"}      & 098.C-0910(A) & A0-G1-J2-J3 & GRAVITY &  22 \& 500 & HD\,55137 \\
    \hline
    2018-11-29\,\tablefootmark{a} & 102.C-0701(B) & A0-G1-J2-K0 & PIONIER & 40 & HD\,51914 \\
    2018-11-30\,\tablefootmark{a} & 102.C-0701(B) & A0-G1-J2-K0 & PIONIER & 40 & HD\,51914 \\
    \hline
    2019-12-15\,\tablefootmark{a} & 104.C-0737(C) & A0-B2-C1-D0 & PIONIER & 40 & HD\,51914 \\
    2019-12-16\,\tablefootmark{a} & 104.C-0737(C) & A0-B2-C1-D0 & PIONIER & 40 & HD\,49741 \\
    2019-12-23\,\tablefootmark{a} & 104.C-0737(A) & D0-G2-J3-K0 & PIONIER & 40 & HD\,51914 \\
    2019-12-24\,\tablefootmark{a} & 104.C-0737(A) & D0-G2-J3-K0 & PIONIER & 40 & HD\,51914 \\
    2019-12-29 & 104.C-0737(B) & A0-G1-J2-J3 & PIONIER & 40 & HD\,51914 \\
    \hline
    2020-11-18\,\tablefootmark{a} & 2020B-M7 & \multicolumn{1}{c|}{W1-S2-S1-E2} & MIRC-X & 102   & HD\,58457 \\ 
    2020-11-19\,\tablefootmark{a} & 2020B-M7 & \multicolumn{1}{c|}{W2-W1-S2-S1} & MIRC-X & 50   & HD\,54930 \\ 
    \hline
    2020-12-13 & 106.21JU.002 & A0-G1-J2-J3 & PIONIER & 40 & HD\,45694, HD\,54438, HD\,51914 \\
    2020-12-19 & 106.21JU.001 & D0-G2-J3-K0 & PIONIER & 40 & HD\,45694, HD\,54438, HD\,51914 \\
    2020-12-28 & 106.21JU.003 & A0-B2-C1-D0 & PIONIER & 40 & HD\,45694, HD\,54438, HD\,51914 \\
    \hline
    \end{tabular}
    \tablefoot{
    \tablefoottext{a}{Observations on consecutive days were grouped into one epoch for continuum analysis.} \\
    }
\end{table*}

\section{Modelling}
\label{sec:modelling}

\subsection{Continuum modelling of the system}
\label{sec:geomodel}

We fit the interferometric visibility and closure phase data at each epoch using the Exeter in-house geometric modelling pipeline \citep[e.g.][]{kreplin18}. As the stellar radii of {MWC\,166\,Aa+Ab} are expected to be \textasciitilde0.04\,mas at the distance calculated by \cite{kaltcheva00}, we assume that the stellar photospheres can be modelled as point sources.

Initially, the visibilities and closure phases of {MWC\,166\,A} were fitted with the following free parameters: separation ($\rho$); position angle\footnote{Defined as East of North} of the secondary component from the primary ($\theta$); and the flux contribution of the secondary to the total flux in the model ($f_2/f_{\mathrm{tot}}$). The primary flux contribution was kept fixed 

Due to the changes in the $f_2/f_{\mathrm{tot}}$ flux ratio, the secondary is at some epochs brighter than the primary in our near-infrared wavelength bands. While there is some variability to the system as a whole over year-length timescales \citep{Pogodin06}, the relative brightness of the two stars has not been previously recorded, so this was an unexpected finding. In light of this, we restricted $\theta$ either to the [$0^{\circ}$, $180^{\circ}$] or [$180^{\circ}$, $360^{\circ}$] range, where the quadrant was chosen for each epoch to achieve an astrometric orbit that is consistent with the spectroscopic orbit of \citet{Pogodin06}. This was done to ensure that the primary and secondary components of the system were correctly identified at each epoch.

Initially, we did not account for contributions from possible dust emission, consistent with the low measured infrared excess emission in the \textit{K}-band \citep{Tjin01}. The 2-point-source model fits allowed us to derive the astrometry of the two components of the system, but yield a flux ratio that changes significantly between epochs, both in the $H$- and $K$-band. These models also consistently overpredicted the visibilities, as can be seen from the red points on Fig.~\ref{fig:vis_mircx_allmodels}, leading to large reduced $\chi^2$ values, in particular on the visibilities (e.g.\ $\chi^2_{\mathrm{vis}}>16$ for the MIRC-X data). In order to reduce this systematic error, we also conducted fits that include extended emission.

\subsubsection{Evidence for extended circumbinary disc emission}
\label{sec:extended_geometry}

We tried modelling the extended flux assuming three different geometries: a Gaussian with full-width at half maximum $\sigma$, seen under inclination $i$ and a major axis (East of North) position angle $\Theta_\mathrm{ext}$; a ring with radius $R$ and a thickness of $0.2R$, seen under inclination $i$ and a position angle $\Theta_\mathrm{ext}$; and as over-resolved flux `background' (modelled as a circular Gaussian with $\sigma=1000$\,mas, i.e.\ filling the field-of-view). The integrated flux contribution of the extended emission component to the total flux in the model is $f_\mathrm{ext}/f_{\mathrm{tot}}$, where we define $f_\mathrm{tot} \equiv f_1 + f_2 + f_\mathrm{ext}$.

For the VLTI epochs, the different geometries for the extended emission returned improved $\chi^2$ values over the 2-point-source model, but no one extended model had consistently smaller $\chi^2$ values over all epochs. Adopting a Gaussian geometry for the extended emission component results in $\chi^2_\mathrm{vis}$ values ranging from 0.34 to 2.24, while adapting a ring geometry results in $\chi^2_\mathrm{vis} = 0.49...4.65$, and overresolved flux results in $\chi^2_\mathrm{vis} = 0.57...6.81$. For comparison, the pure point-source model has $\chi^2_\mathrm{vis}$ between 0.94 and 31.86. Closure phase $\chi^2$ values were found to be almost completely model-independent, with the models returning $\chi^2_\mathrm{CP}~<~2.84$ (point-sources), $\chi^2_\mathrm{CP}~<~1.62$ (background), $\chi^2_\mathrm{CP}~<~1.13$ (Gaussian), $\chi^2_\mathrm{CP}~<~1.43$ (ring).

\begin{figure}
    \centering
    \includegraphics[width=\columnwidth]{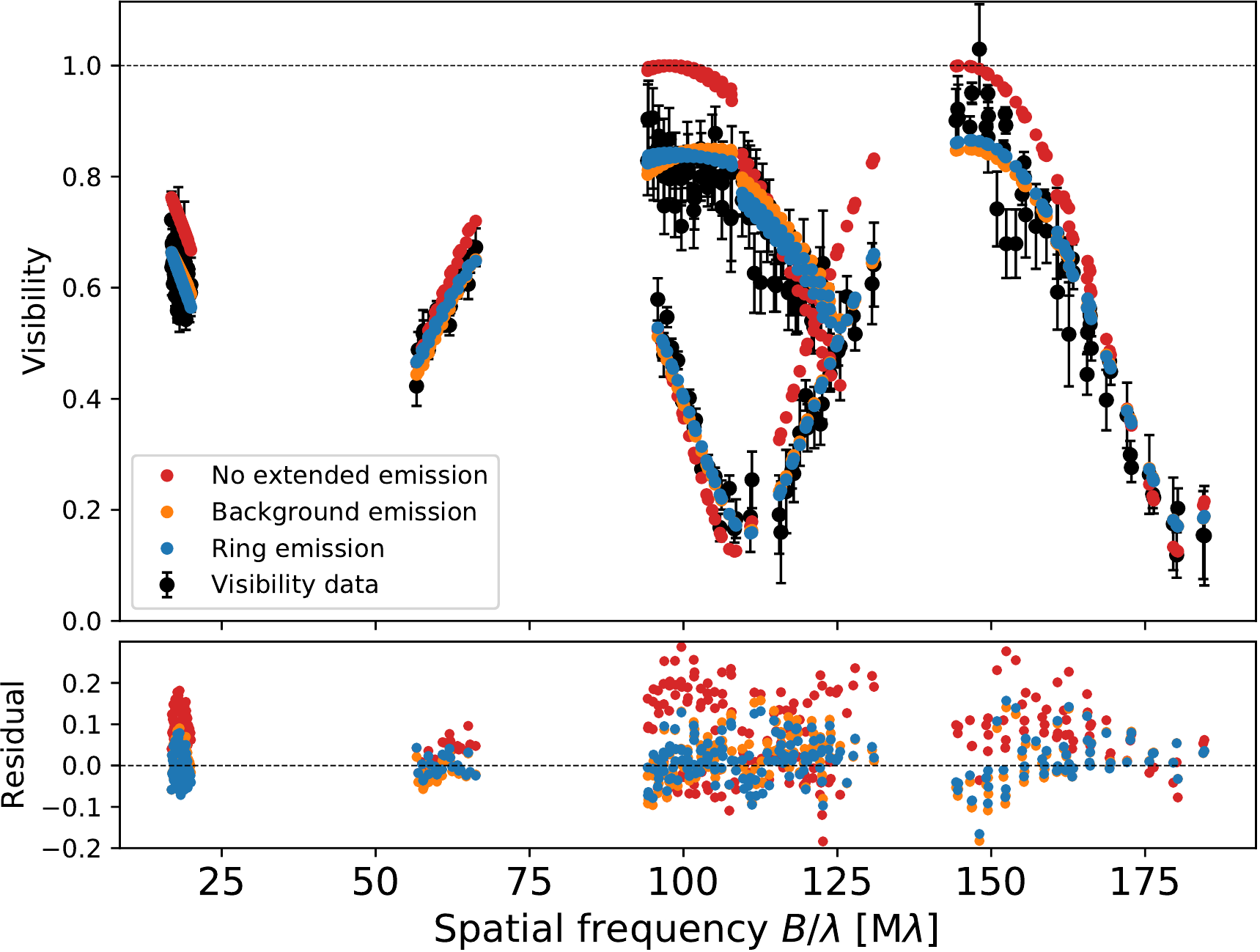}
    \caption{Visibilities (and associated residuals) of MIRC-X models. At $V\gtrsim0.6$, the purely point-source model overshoots the observed datapoints substantially.}
    \label{fig:vis_mircx_allmodels}
\end{figure}

\begin{figure}
    \centering
    \includegraphics[width=\columnwidth]{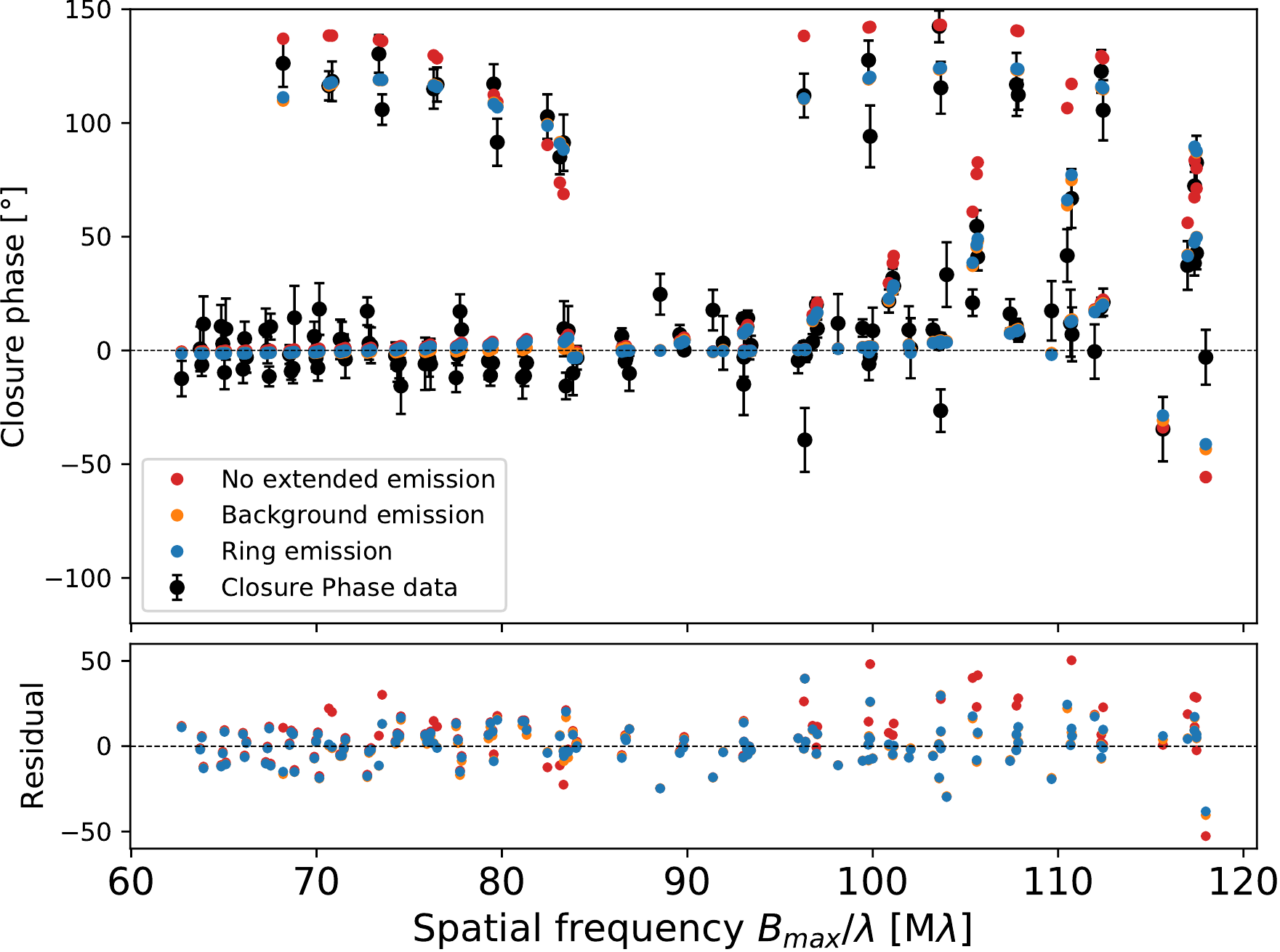}
    \caption{Closure phases (and associated residuals) of MIRC-X models.}
    \label{fig:cp_mircx_allmodels}
\end{figure}

The MIRC-X data probes {\textasciitilde$3\times$} higher spatial frequencies than the VLTI data. Data taken over a larger range of spatial frequencies allows us to probe further lobes of the visibility curve, which helps to more reliably distinguish the effects of the extended emission from the sinusoidal binary modulation. Figures \ref{fig:vis_mircx_allmodels} and \ref{fig:cp_mircx_allmodels} respectively show the resultant model visibilities and closure phases obtained from our geometric models described above, overlaid over the data. While the closure phases are well-described by the original 2-point-source model independently of any extended flux or lack thereof, for the visibilities this is not the case. The 2-point-source model, represented by the red points on Fig. \ref{fig:vis_mircx_allmodels}, clearly overpredicts the visibility compared to the other models, especially in the high-visibility regime (where $V\gtrsim0.6$). The model parameters corresponding to Figs.~\ref{fig:vis_mircx_allmodels} and \ref{fig:cp_mircx_allmodels} are shown in Table~\ref{tab:MIRCXmodels}.

By examining the $\chi^2$ values both for visibility and closure phase in Table~\ref{tab:MIRCXmodels}, it can be seen that the background model provides a significant improvement on the 2-point-source model. A ring profile provides a similar, or even slightly better fit (see Table~\ref{tab:MIRCXmodels}), but introduces 3 additional free parameters while providing only a marginal improvement in the goodness of the fit. We also conducted a fit for a Gaussian model ($\sigma, i, \Theta_\mathrm{ext}$) which returned similar $\chi^2$ values to the ring model, but we found that the parameters $i$ and $\Theta_\mathrm{ext}$ did not converge to a value independent of the boundary conditions chosen, while the flux parameters $f_2/f_{\mathrm{tot}}$ and $f_\mathrm{ext}/f_{\mathrm{tot}}$ were found to be consistent with the background model. As such, we favour the background model over the Gaussian model.

\begin{figure}
    \centering
    \includegraphics[width=\columnwidth]{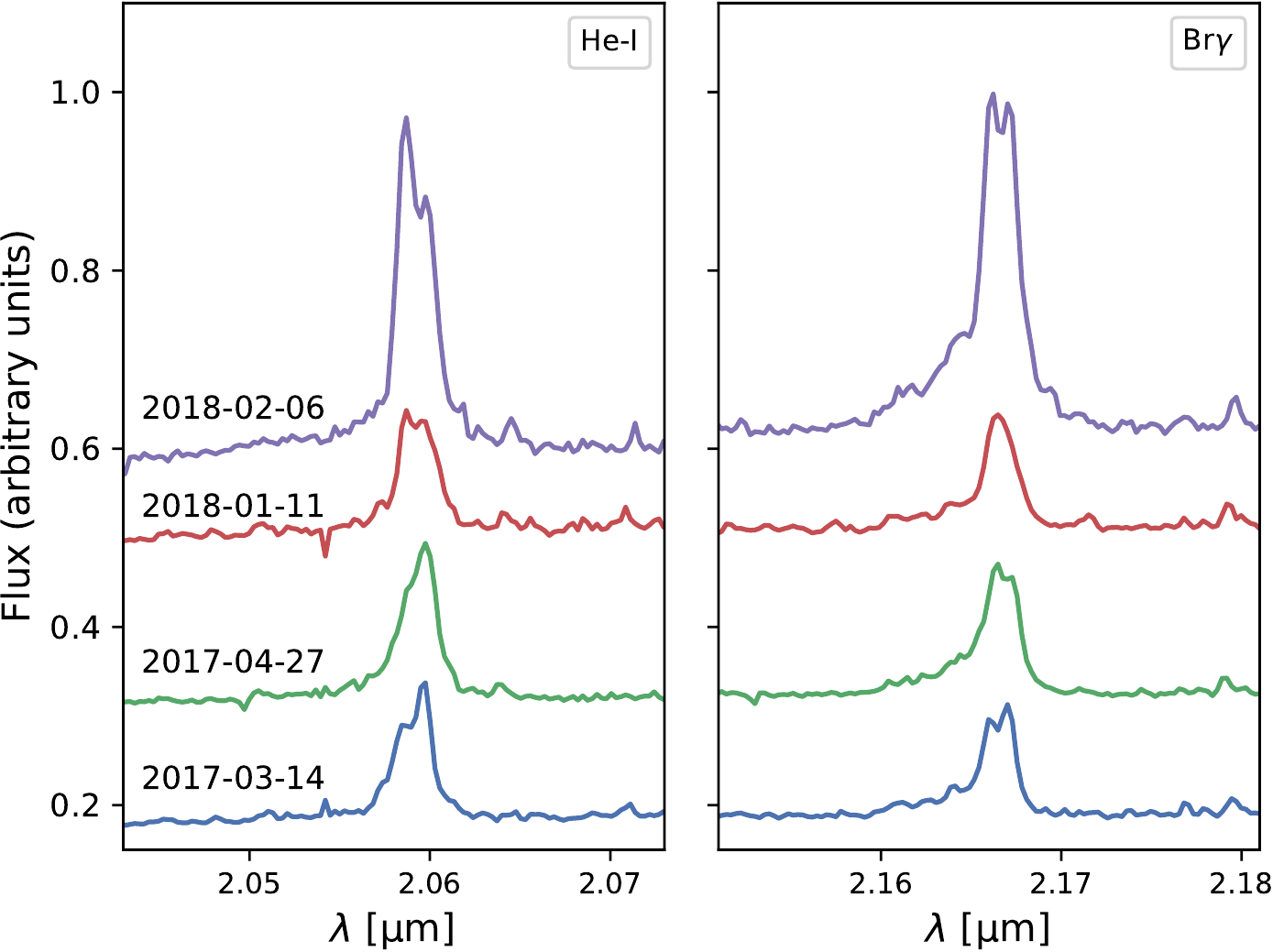}
    \caption{Continuum-normalised spectra around the He\,\textsc{i} and Br$\gamma$ lines. The different epochs have been offset for clarity. Epoch-dependent variations are visible.}
    \label{fig:flux}
\end{figure}

\begin{table}
    \caption{MIRC-X extended emission model comparison.}
    \centering
     \begin{tabular}{l | l | l | S{l}}
     \hline
     Model: & No ext.\ emission & Background & Ring \\
     \hline
     $\rho$ [mas] & $2.87\pm0.01$ & $2.84\pm0.01$ & $2.84\pm0.01$\\
     $\theta$ [\textdegree] & $134.3\pm0.4$ & $131.1\pm0.3$ & $131.7\pm0.3$ \\
     $f_2/f_{\mathrm{tot}}$ & $0.437\pm0.012$ & $0.345\pm0.004$ & $0.344\pm0.006$ \\
     \hline
     $R$ [mas] & - & - & $4.55^{+0.22}_{-0.19}$ \\
     $i$ [\textdegree] & - & - & $55.2\pm4.0$ \\
     $\Theta_\mathrm{ext}$ [\textdegree] & - & - & $100.6^{+13.1}_{-9.8}$ \\
     $f_\mathrm{ext}/f_{\mathrm{tot}}$ & - & $0.150\pm0.004$ & $0.150\pm0.005$\\
     \hline
     $\chi^2_\mathrm{Vis}$ & 16.47 & 3.27  & 2.63 \\
     $\chi^2_\mathrm{CP}$ & 4.28 & 1.95 & 2.07 \\
     \hline
     \end{tabular}
    \label{tab:MIRCXmodels}
\end{table}

Therefore, we adopt the overresolved background model as geometry for the extended emission for all epochs and instruments, likely representing scattered light from the disc. This minimised the model complexity and degrees of freedom. The relative astrometry was therefore fitted with four free parameters: separation ($\rho$); position angle of the secondary component from the primary ($\theta$); secondary flux as fraction of the total flux ($f_2/f_{\mathrm{tot}}$) and extended flux as fraction of the total flux ($f_\mathrm{ext}/f_{\mathrm{tot}}$).

It is apparent that the relative astrometry of the binary ($\rho, \theta$) does not depend much on whether extended flux is included in the fit. The value of $f_2/f_{\mathrm{tot}}$ and $f_\mathrm{ext}/f_{\mathrm{tot}}$ change depended on whether extended emission is included in the fit, but is rather independent of the geometry of the emission. 

\begin{figure*}
	\includegraphics[width=\textwidth]{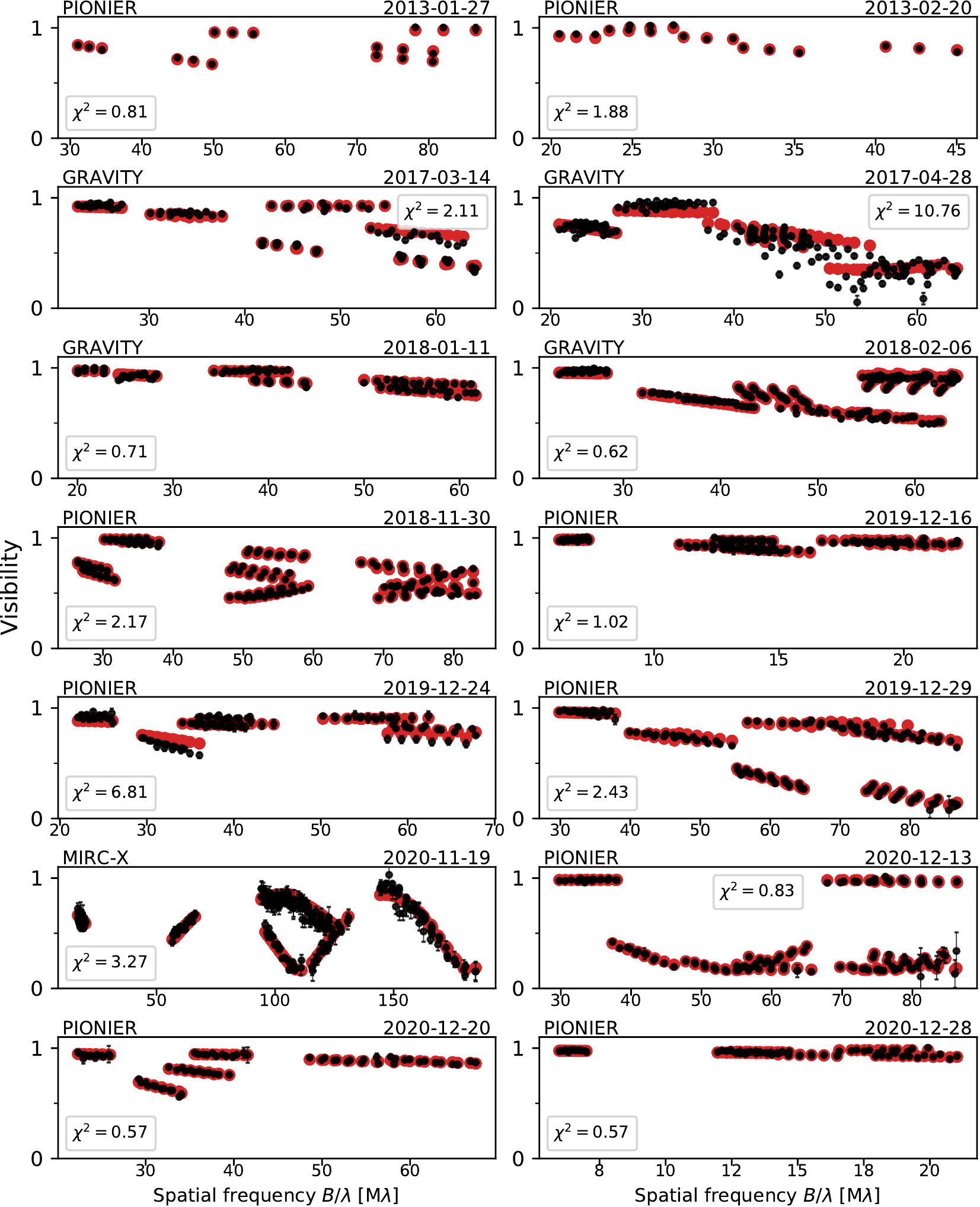}
	\caption{Observed continuum visibilities (black) and corresponding models (red) plotted against spatial frequency for all epochs. The model fit included extended background emission.}
    \label{fig:vis_all}
\end{figure*}

\begin{figure*}
	\includegraphics[width=\textwidth]{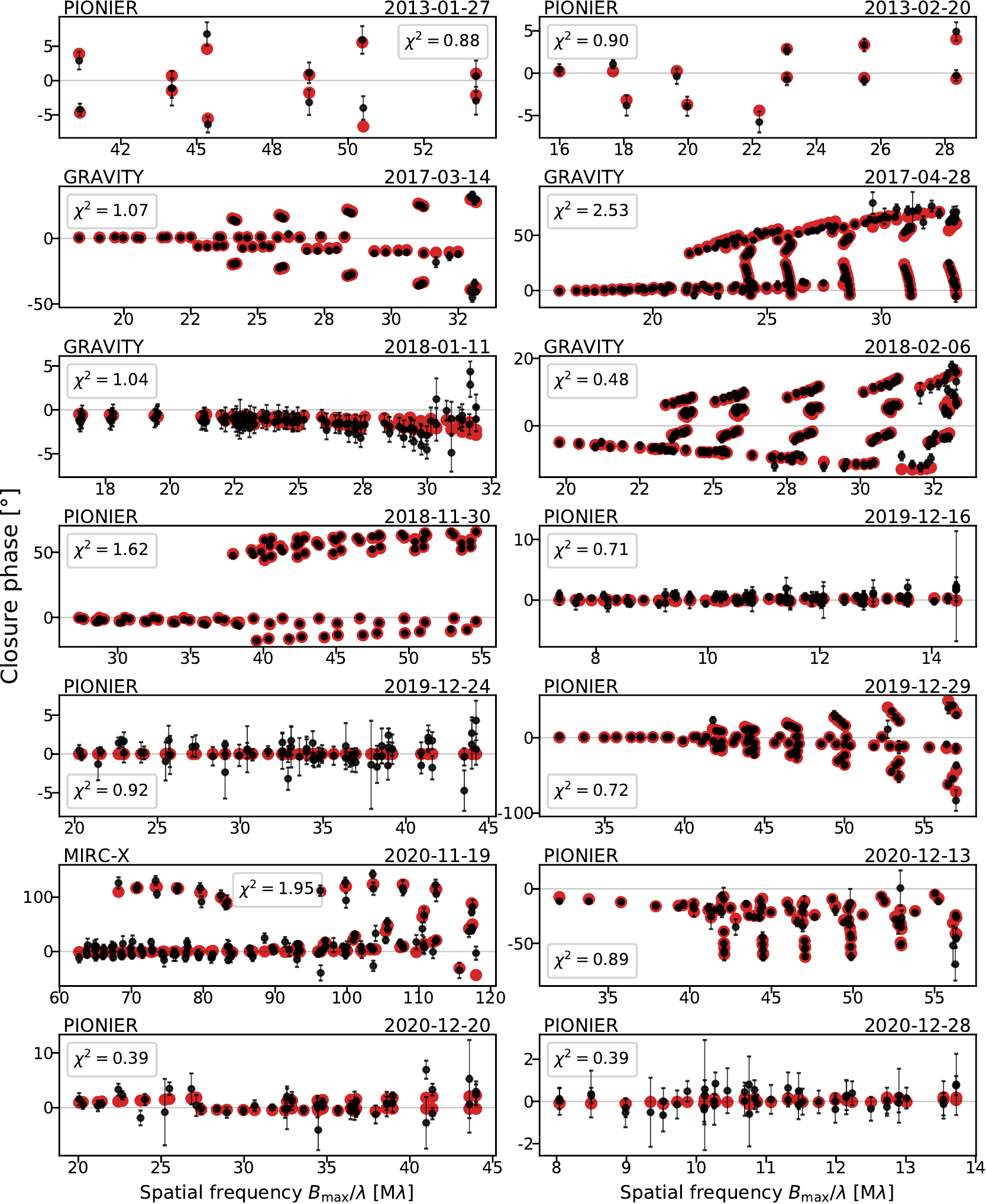}
	\caption{Observed continuum closure phases (black) and corresponding models (red) plotted against spatial frequency for all epochs. The model fit included extended background emission. The scaling on the vertical axis was adjusted for each epoch.}
    \label{fig:cp_allepochs_sf}
\end{figure*}

\begin{figure}
	\includegraphics[width=\columnwidth]{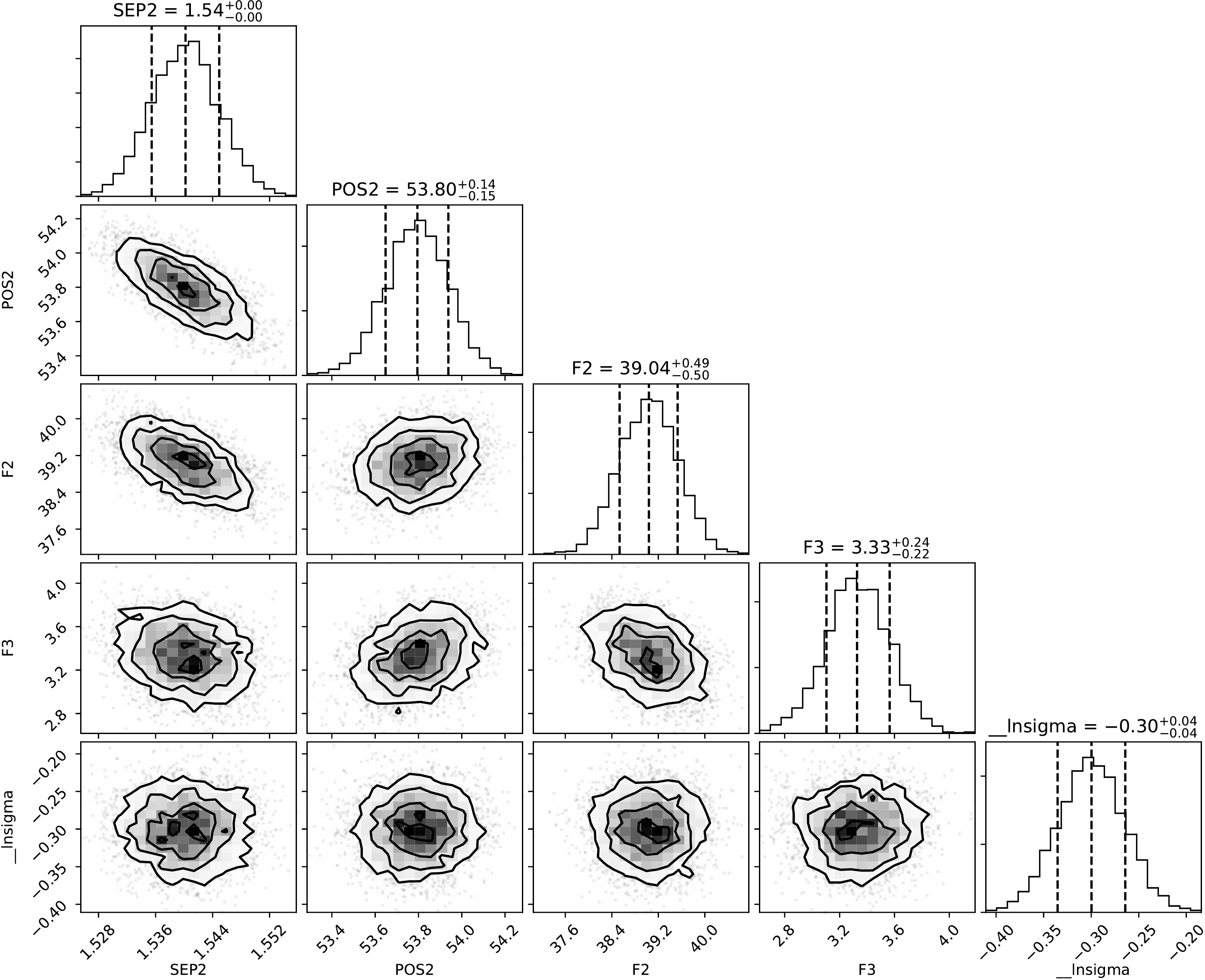}
	\caption{Corner plot showing the possible correlations between free parameters ($\rho$, $\theta$, $F_2=100\cdot f_2/f_{\mathrm{tot}}$ and $F_3=100\cdot f_\mathrm{ext}/f_{\mathrm{tot}}$ respectively) for epoch 2018-02-06 of the continuum GRAVITY data, where the model fit included extended background emission.}
    \label{fig:cornerplot}
\end{figure}

\begin{table*}
    \caption{Relative astrometry for MWC\,166 Aa+Ab, derived from $H$- and $K$-band continuum visibility and closure phase modelling. The model included two point-sources and extended background emission, for a total of four free parameters: separation $\rho$, position of secondary component $\theta$ (East of North), secondary flux $f_2/f_{\mathrm{tot}}$, and extended flux $f_\mathrm{ext}/f_{\mathrm{tot}}$.}
    \centering
    \begin{tabular}{l | c | S{l} S{l} | S{l} S{l} | S{c} S{c}}
    \hline
    Epoch & Inst. & $\rho$ [mas] & $\theta$ [\textdegree] & $f_2/f_{\mathrm{tot}}$ & $f_\mathrm{ext}/f_{\mathrm{tot}}$ & $\chi^2_\mathrm{vis}$ & $\chi^2_\mathrm{CP}$\\
        \hline
    2013-01-27 & \texttt{PIONIER} & $1.50\pm0.09$ & $309.8\pm0.3$ & $0.223\pm0.015$ & $0.021\pm0.005$ & 0.81 & 0.88\\
    2013-02-20 & " & $1.68\pm0.12$ & $343.5\pm1.1$ & $0.189\pm0.021$ & $0.002\pm0.001$ & 1.88 & 0.90\\
    \hline
    2017-03-14  & \texttt{GRAVITY} & $1.59\pm0.08$ & $358.7\pm0.4$ & $0.289\pm0.003$ & $0.051\pm0.004$ & 2.11 & 1.07\\
    2017-04-28\tablefootmark{a}  & " &  $1.90\pm0.10$ & $47.4\pm0.2$ & $0.268\pm0.003$ & $0.116\pm0.005$ & 10.76 & 2.53\\
    2018-01-11  & " & $0.89\pm0.05$ & $233.3\pm1.0$ & $0.295\pm0.013$ & $0.015\pm0.003$ & 0.71 & 1.04\\
    2018-02-06  & " & $1.54\pm0.08$ & $306.2\pm0.2$ & $0.274\pm0.003$ & $0.023\pm0.002$ & 0.62 & 0.48\\
    \hline
    2018-11-30\,\tablefootmark{a} & \texttt{PIONIER} & $2.43\pm0.12$ & $138.2\pm0.1$ & $0.271\pm0.001$ & $0.013\pm0.002$ & 2.17 & 1.62\\
    2019-12-16\,\tablefootmark{a} & " & $2.03\pm0.11$ & $148.1\pm0.8$ & $0.508\pm0.005$ & $0.011\pm0.002$ & 1.02 & 0.71\\
    2019-12-24\,\tablefootmark{a} & " & $1.34\pm0.07$ & $154.2\pm0.6$ & $0.456\pm0.002$ & $0.087\pm0.004$ & 6.81 & 0.92\\
    2019-12-29 & " & $1.34\pm0.07$ & $167.5\pm0.1$ & $0.429\pm0.001$ & $0.023\pm0.002$ & 2.43 & 0.72 \\
    \hline
    2020-11-19\,\tablefootmark{a} & \texttt{MIRC-X} & $2.84\pm0.14$ & $131.1\pm0.3$ & $0.345\pm0.004$ & $0.150\pm0.004$ & 3.27 & 1.95 \\
    \hline
    2020-12-13 & \texttt{PIONIER} & $2.09\pm0.10$ & $144.6\pm0.1$ & $0.410\pm0.001$ & $0.013\pm0.002$ & 0.86 & 0.89 \\
    2020-12-19 & " & $1.85\pm0.09$ & $149.7\pm0.2$ & $0.422\pm0.008$ & $0.035\pm0.002$ & 0.72 & 1.03 \\
    2020-12-28 & " & $1.53\pm0.11$ & $162.4\pm1.3$ & $0.675\pm0.058$ & $0.018\pm0.005$ & 0.57 & 0.39 \\
    \hline
    \end{tabular}
    \label{tab:bg_models}
    \tablefoot{
    \tablefoottext{a}{Data combines two days of observation modelled simultaneously, as defined in table \ref{tab:MWCobs}.}
    }
\end{table*}

\subsection{Modelling of the $K$-band He\,\textsc{i} and Br\,$\gamma$ lines}
\label{sec:linemodelling}

For the GRAVITY data, we simultaneously recorded high-resolution data ($\Delta\lambda/\lambda~=~4000$) for all epochs, in addition to the low-resolution data used to establish the relative astrometry. As can be seen from MWC\,166\,A's $K$-band spectrum, there are strong line features at 2.058 and $2.166\,\upmu\mathrm{m}$ (Fig.~\ref{fig:flux}), corresponding to Helium-\textsc{i} and Brackett-$\gamma$ emission respectively.

In order to model the spectral lines, we used the fitting tool \texttt{PMOIRED}\footnote{Reference: \url{https://github.com/amerand/PMOIRED}}, which is capable of fitting closure phases, differential phases, differential visibilities, and line spectra simultaneously, both over the continuum and over specific spectral windows.

After finding the continuum geometry of the system for each epoch (Sect.~\ref{sec:extended_geometry}), we introduced new model components to fit the He\,\textsc{i} and Br\,$\gamma$ lines individually, using geometries of varying complexity. We initially used a single Gaussian of full-width-at-half-maximum (FWHM) $\sigma=0.1$\,mas and left the position of the line-emitting region as a free parameter. At all epochs, this resulted in only very small spatial displacements from the origin, suggesting the primary component is responsible for the majority of the emission in the system. However, this singular emission zone near the primary is perhaps too simplistic a model. If we examine the flux intensity of the two lines (Fig.~\ref{fig:flux}), there are indications of time-dependent variability, as well as signs of double-peaked lines at several epochs. This could imply that the emission originates from both stars, or it could a signature of the gas kinematics.  Furthermore, the differential phases over the spectral lines show signatures of rotating emission which are not accurately modelled by the single-Gaussian model. These features would be most naturally explained by a circumprimary disc.

\subsubsection{Circumprimary gas disc model}
\label{sec:PMOIRED_model}

The circumprimary disc model we used is based on the model described in \citet{frost22}, which was used to model the binary system {HR\,6819}. The model is comprised of several components, and has 12 free parameters which describe the entirety of the system.

Firstly, the stars themselves are modelled as uniform discs with diameters corresponding to twice the stellar radii we obtained from our continuum fit (see Sect.~\ref{sec:stellarproperties}), and the secondary component is given a displacement from the primary's position at the origin, as well as a continuum flux $f_2/f_1$, while the primary flux is fixed to $f_1\equiv1$.

The line emission is subsequently modelled to be originating from two regions, one blue-shifted and the other red-shifted, to represent the approaching and receding part of a rotating disc (labelled $B$ and $R$ respectively). Each of these components was given its own spatial displacement $(x_i, y_i)$, with the size of the emitting region following a Gaussian profile with FWHM $\sigma_i = \frac{1}{2}\sqrt{x^2_i + y^2_i}$. We additionally modelled the line components in the spectral domain. Each component is given a flux profile $F_i = f_i + F_\mathrm{L}$, consisting of a Lorentzian component $F_\mathrm{L}$ which was kept equal for both wings, and a flat component $f_i$, accounting for the differences in line strength between the components (which can be seen to vary by epoch in Fig.~\ref{fig:flux}). Each line wing is centred on a wavelength which is displaced from the central line wavelength $\lambda_0$, such that $\lambda_\mathrm{B}=(\lambda_0-\Delta\lambda)$ and $\lambda_\mathrm{R}=(\lambda_0+\Delta\lambda)$.

The resultant fitted parameters of the model described above are presented for all GRAVITY epochs in Sect.~\ref{sec:PMOIRED_results}. The model is fitted simultaneously to the telluric-corrected spectrum, closure phase, differential phase, and differential visibility.

\section{Results: Orbital solution and mass/distance constraints}
\label{sec:results}

\subsection{Binary astrometry and orbital fit}
\label{sec:astrometry}

The parameters of our best-fit continuum model with background component are listed in Table~\ref{tab:bg_models}. The model visibilities that correspond to the best-fit model are shown in Fig.~\ref{fig:vis_all}, while closure phases are shown in Fig.~\ref{fig:cp_allepochs_sf}. Our modelling script makes use of Markov Chain Monte Carlo module \texttt{emcee} \citep{corner} to explore the parameter space and obtain error estimates from the posterior probability distribution. We show the corner plot for a representative epoch (2018-02-06) in Fig.~\ref{fig:cornerplot}. An important source of systematic uncertainty that affects primarily the derived separations is the wavelength calibration, and we account for this by including a systematic uncertainty of 5\% \citep{gallenne18} for the separations listed in Table~\ref{tab:bg_models}.

Using the relative astrometry for each epoch, we fitted a Keplerian orbit using the standard Campbell elements: {$P=$ orbital} period; $T_0~=$ epoch of periastron passage; $a_1=$ semi-major axis of primary component; $i=$ orbital inclination (to line of sight); $e =$ eccentricity; $\Omega =$ longitude of ascending node; $\omega =$ longitude of periastron; $K_1 =$ orbital curve semi-amplitude of primary component; $V_0 =$ RV of the system's centre of mass.

We used two fitting approaches:

\textit{\bf \texttt{ORBITX code:}} This code\footnote{Available at: \url{https://zenodo.org/record/61119}} \citep{tokovinin_orbitx} fits orbits using both astrometric and RV data simultaneously. We modified the \texttt{ORBITX} code to account for uncertainties in both $\rho$ and $\theta$, a feature absent from the original code, which only uses uncertainties on $\rho$.

\textit{\bf Grid-search algorithm:} We used the grid-search algorithm developed by \citet{kraus09_th1oric} to construct a grid of orbital solutions in the $P$, $T_0$, and $e$ parameter space, where the remaining elements $a, i, \Omega, \omega$ are determined from the Thiele-Innes elements. We explored the parameter space around $P = 0.480 ... 1.100\,\mathrm{yr}$, $T_0 = 2019.5 ... 2020.5\,\mathrm{yr}$ (step sizes of 0.001\,yr), and $e = 0.100 ... 0.600$ with a step size of 0.001, and selected the solution with the lowest combined residuals in RV and astrometry. We then repeated the process with smaller step-sizes around the initial solution (a factor of ten for all parameters, $=0.0001\,\mathrm{yr}, 0.0001\,\mathrm{yr}, 0.0001$ respectively for $P, T, e$), in order to increase precision. Uncertainties were calculated by examining the $\chi^2$ curve for each parameter.

\begin{figure}
    \centering
    \includegraphics[width=\columnwidth]{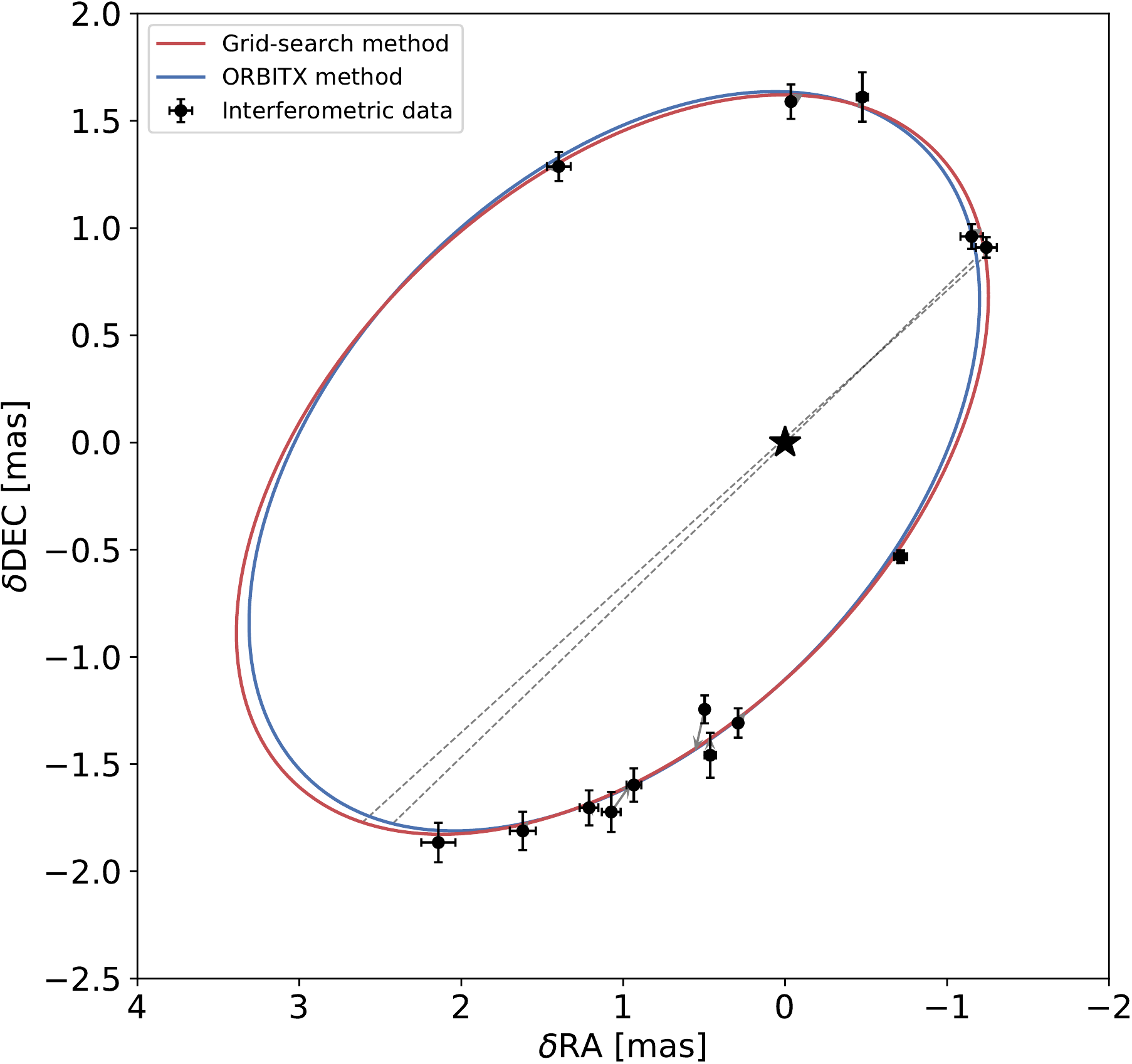}
    \caption{Astrometric orbit solutions derived using both the \texttt{ORBITX} code (blue line) and grid-search code (red line). The primary star is kept fixed at the origin, and the $x$- and $y$-axes show displacement in right ascension and declination respectively. The dotted lines connect the ascending and descending nodes of each orbit.}
    \label{fig:posplot}
\end{figure}

\begin{figure}
    \centering
    \includegraphics[width=\columnwidth]{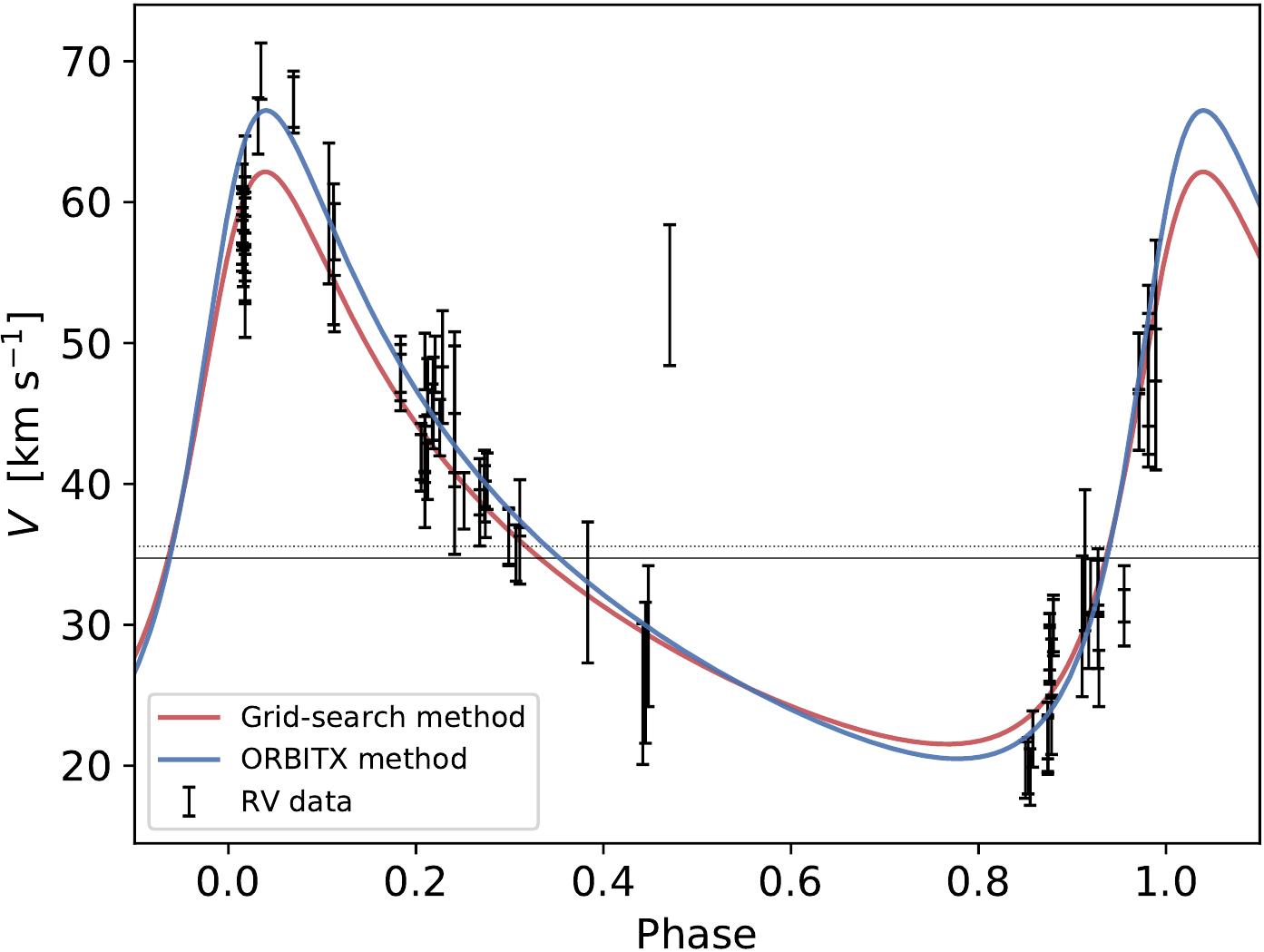}
    \caption{Radial velocity measurements of MWC\,166\,A taken in the period 1994--2005 plotted against orbital phase. The blue and red fitted curves correspond to the orbits specified in columns (3) and (4) of table \ref{tab:orbits}, respectively. The black dotted line shows the velocity of the system's centre of mass ($V_0$) for the grid-search method, and the solid grey line shows $V_0$ for the \texttt{ORBITX} orbit.}
    \label{fig:RVplot}
\end{figure}

The best-fit orbit solutions found with these methods are listed in columns (3), (4) of Table~\ref{tab:orbits}, and are overplotted on the data in Figs.~\ref{fig:posplot} and \ref{fig:RVplot}. Both the orbits provide a very good fit to the existing data, and provide similar results for all parameters. This is despite a substantial portion of the orbit still lacking astrometric observations. We adopt the orbit from column (3) when discussing derived quantities in the subsequent sections.

\subsection{Comparison to RV orbit}

\citet{Corporon99} and \citet{Pogodin06} derived orbits for MWC\,166\,A from the RV data, with the more recent of the two being a refinement including additional RV points. The orbital parameters for this RV orbit are shown in column (2) of Table\,~\ref{tab:orbits}. Our spectroscopic+astrometric orbital solutions differ substantially from the earlier RV-only orbit. The most notable difference is in the orbital period, which we calculated as almost exactly twice the length of \citet{Pogodin06}'s orbit. This doubling of the period was only discernible thanks to our astrometric data, as using the radial velocities alone provides an equally good fit to both orbits. We also found the orbit to be much more elliptical than previously thought, with its eccentricity of $0.498\pm0.003$ being much larger than that of the RV orbit ($e~=~0.28\pm0.03$). A newly determined parameter from our orbit is the inclination, with a value of $i~=~53.6\pm0.3^\circ$.

\begin{table*}
    \caption{Orbital parameters for MWC\,166\,A. Column (2) gives the RV fit obtained by \citet{Pogodin06}. Columns (3) and (4) give the best-fit orbital solution including both RV data and the astrometry data, using the \texttt{ORBITX} and the grid-search methods respectively. These solutions were derived using the background geometric model described in Sect.~\ref{sec:extended_geometry}.}
    \centering
    \begin{tabular}{S{l} | c | c c }
        \hline
        Parameter (1) & RV only (2) & \texttt{ORBITX} orbit (3) & Grid-search orbit (4) \\ [0.5ex]
        \hline
        $P$ [yr] &  $0.50296\pm0.00027$ & $1.0067\pm0.0001$ & $1.0066\pm0.0002$ \\
        $P$ [days] &  $\mathit{183.70\pm0.10}$ & $\mathit{367.69\pm0.04}$ & $\mathit{367.65\pm0.07}$ \\
        $T_0$ [yr] & $1993.3581\pm0.0078$  & $2020.0722\pm0.0003$ & $2020.0713\pm0.0010$\\
        $a_1$ [mas] & - & $2.6122\pm0.0385$ & $2.684\pm0.008$\\
        $i$ [$^{\circ}$] & - & $53.62\pm0.32$ & $55.27\pm0.11$ \\
        $a_1\sin{i}$ [mas] & $2.15\pm0.07$\tablefootmark{a} & - & -\\
        $e$ & $0.28\pm0.03$ & $0.498\pm0.001$ & $0.492\pm0.003$ \\
        $\Omega$ [$^{\circ}$] & - & $306.3\pm0.2$ & $304.9\pm0.2$ \\
        $\omega$ [$^{\circ}$] & $263.8\pm6.6$ & $313.8\pm0.2$ & $315.6\pm0.5$ \\
        \hline
        $K_1$ $\mathrm{[km~s^{-1}]}$ & $18.6\pm0.7$ & $20.3\pm0.4$ & $20.3\pm0.4$\\
        $V_0$ $\mathrm{[km~s^{-1}]}$ & $44.2\pm0.5$ & $35.3\pm0.3$ & $34.7\pm0.3$ \\
        $M_\mathrm{tot}$ [$M_\odot$]\tablefootmark{b} & - & $17.05\pm2.70$ & $18.52\pm2.81$\\
        \hline
    \end{tabular}
    \label{tab:orbits}
    \tablefoot{
    \tablefoottext{a}{\citet{Pogodin06} returns $a\sin{i} = 60\pm2~ R_\odot$. A conversion to milliarcseconds has been made to allow better comparison with the calculated semi-major axis and inclination. The distance used was 990\,pc.\\}
    \tablefoottext{b}{Mass calculated for $d = (990\pm50)$\,pc.}
    }
\end{table*}

\subsection{Dynamical system mass}
\label{sec:kepler}

According to Kepler's third law, the period of an orbit $P$ is proportional to the cube of the semi-major axis of the orbit $a$. Using the usual angular diameter-distance relation ${a~\mathrm{[au]} = a~\mathrm{[\arcsec]}\times d~\mathrm{[pc]}}$, we can show the dependence of the total system mass $M_\mathrm{tot} \equiv (M_1 + M_2)$ (in solar masses) on distance $d$ (in parsecs), where $G$ is the gravitational constant and $M_{1,2}$ the masses of the two stars:
\begin{equation}
M_\mathrm{tot} = \frac{4 \pi a^3 \mathcal{Z}}{G P^2}~d^3,
\label{eqn:CombinedMass}
\end{equation}
where $\mathcal{Z} = 1684.14~\mathrm{m^3}{M_\odot}^{-1}$ is a constant introduced to account for the change in units from au to metre, and from kg to $M_\odot$. In the above equation, the total mass is also known as the `dynamical mass', signifying it is derived from fitting the dynamical orbit of the system. Since $M_\mathrm{tot} \propto d^3$, a reliable distance value is needed to obtain rigorous mass estimates.

Unfortunately, there are several conflicting distance estimates for this system in the literature. As mentioned in Sect.\,\ref{sec:intro}, photometrically calculated distances have placed MWC\,166 at a distance of {\textasciitilde$1\,\mathrm{kpc}$}. Parallax observations have corroborated the photometric distances for a majority of the other individual members of CMa\,OB1, but are not consistent in the case of MWC\,166. Hipparcos \citep{HIP97} measured a distance of $247\pm82$\,pc, while Gaia Data Release 2 (DR2) parallaxes correspond to an even shorter distance of $131^{+16}_{-13}$\,pc \citep{BailerJones18} -- roughly ten times closer than the {\textasciitilde1\,kpc} to its parent association. This discrepancy can likely be explained by the binarity of the system, as DR2 does not solve for source multiplicity. Indeed, the recent release of preliminary results from Gaia EDR3 has brought the parallax distance to MWC\,166 closer to the photometric distance values, albeit with very large uncertainties ({$d\sim1600\pm700$\,pc}, \citealt{bailerjones20}). Our Keplerian mass-distance relation (Eq.~\ref{eqn:CombinedMass}) shows that distances of $\lesssim 650$\,pc correspond to masses of $M_\mathrm{tot} < 5\,M_\odot$, clearly below the mass threshold for MWC\,166\,Aa's spectral type of B0III \citep{Fairlamb15}. Conversely, the photometric distances offer more physically realistic values -- the most recent distance estimate of ${990\pm50~\mathrm{pc}}$ returns a system mass of $M_\mathrm{tot} = {17.1\pm2.7~M_\odot}$, which is also in agreement with the prediction of 20-25 $M_\odot$ by \citet{Pogodin06}. Additionally, any parallax distances are unreliable due to having been calculated with the assumption of a six-month orbit, which we have shown is too short by a factor of two. In light of these points, in this work we have treated the photometric distances preferentially. We have therefore taken the literature distance to MWC\,166 to be the most recent photometric distance, $990\pm50$\,pc \citep{kaltcheva00}.

\subsection{From combined mass to individual masses and other properties}
\label{sec:stellarproperties}

Our full orbital solution allows us to constrain model-independent individual masses for the first time. If the system's orbital period ($P$), eccentricity ($e$) and inclination ($i$) are known, as well as the RV semi-amplitude of the primary component ($K_1$), it is possible to calculate the binary mass function $f$ \citep[e.g.][]{Boffin12, Cure15}:

\begin{equation}
    f \equiv \frac{K_1^3 P\, \big(1 - e^2\big)^{3/2}}{2\pi G} = \frac{\left(M_2 \sin i\right)^3}{M_\mathrm{tot}^2},
    \label{eqn:binarymassfunction}
\end{equation}
which equates the mass of the secondary component to the other elements. Rearranging for $M_{2}$ gives:
\begin{equation} 
    M_2 = \frac{1}{\sin i}~\Bigg[ \frac{K_1^3 P\, \big(1 - e^2\big)^{3/2}}{2\pi G}\cdot M_\mathrm{tot}^2\Bigg]^{1/3},
    \label{eqn:m2fromorbit}
\end{equation}
and the mass of the primary can therefore be trivially found through $M_1~=~M_\mathrm{tot}~-~M_2$.

Using this method we determined the masses of MWC\,166\,Aa and MWC\,166\,Ab as $M_1=(12.19~\pm~2.18)\,M_\odot$ and $M_2=(4.90~\pm~0.52)\,M_\odot$, respectively. Our $M_{\mathrm{dyn}}$ measurements for both components were used to derive the remaining stellar parameters, and their respective confidence bounds, from theoretical evolution tracks. 
We used CMD 3.6\footnote{\url{http://stev.oapd.inaf.it/cgi-bin/cmd}} to generate a Solar metallicity ($Z=0.0152$) isochrone table from PARSEC 1.2S \citep{bressan_parsec_2012, chen_improving_2014, chen_parsec_2015, tang_new_2014, marigo_new_2017, pastorelli_constraining_2019}.

\begin{figure}
    \centering
    \includegraphics[width=\columnwidth]{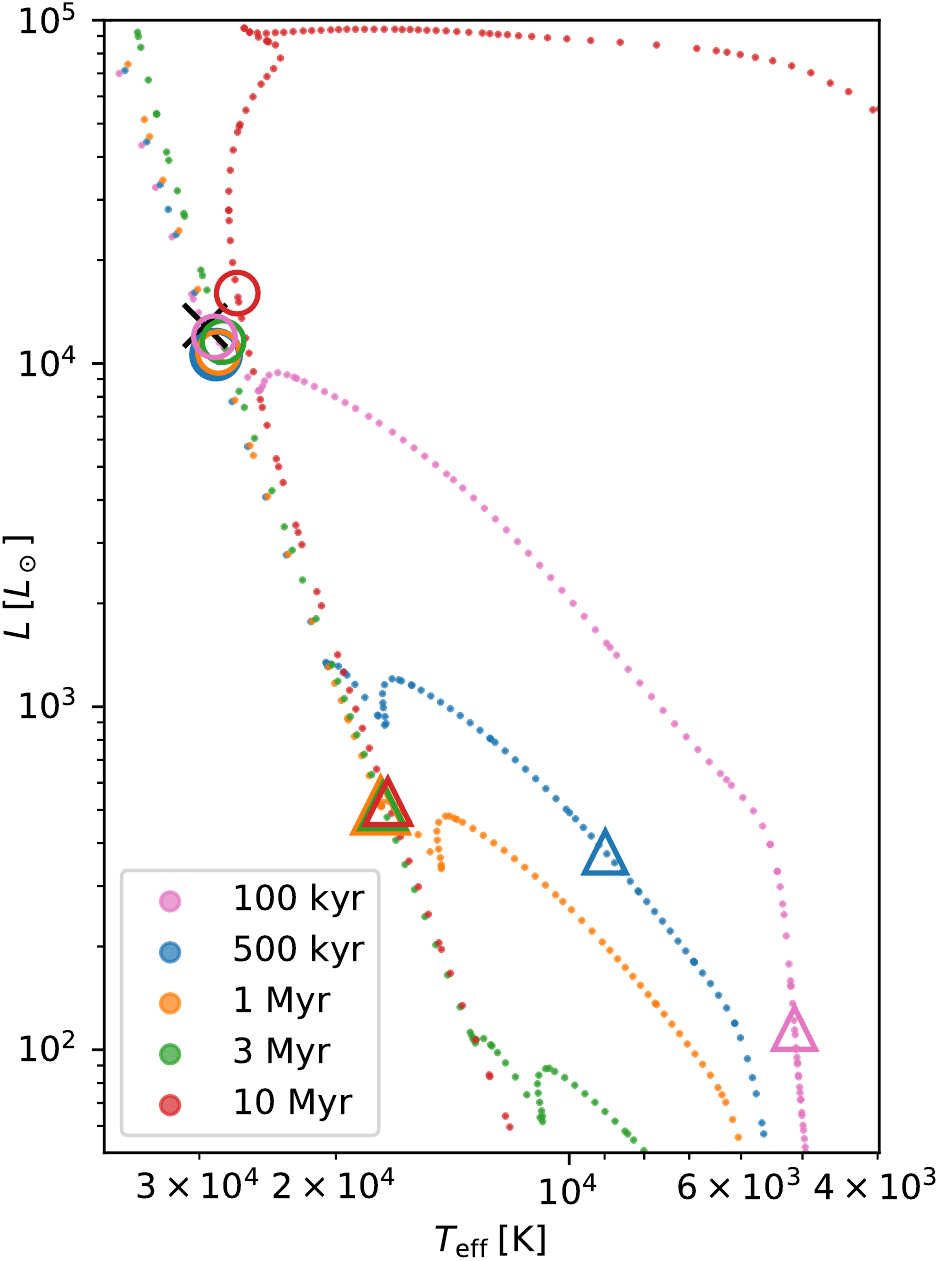}
    \caption{HR diagram showing PARSEC 1.2S isochrone tracks for each age. Superimposed are the interpolated $L$, $T_\mathrm{eff}$ values for the primary star (large circles) and secondary star (triangles), for all ages. The black cross shows the location of the primary's parameters as determined by \citet{Fairlamb15}.}
    \label{fig:hr_diagram}
\end{figure}

The evolutionary status of MWC\,166\,A has not been conclusively established. Analysis of the spatial distribution of O- and B-type stars in its parent OB association, CMa\,OB1/R1, resulted in an estimate for its age of {\textasciitilde$3\times10^6$\,yr} \citep{Claria74a}. At this age, MWC\,166\,Aa is likely already onto the main sequence, while MWC\,166\,Ab might still be in its pre-main-sequence stage \citep{Tjin01}. Later work by \citet{Herbst77} showed that CMa\,OB1/R1's main stars are located on the rim of an expanding shell of neutral hydrogen, consistent with star formation being triggered by a supernova within the last \textasciitilde500\,kyr. Due to the uncertain age of the system, we selected these two potential ages, as well as a reasonable intermediate age ($1\times10^6$\,yr), upper bound ($1\times10^7$\,yr), and lower bound ($1\times10^5$\,yr). Then, for each of the estimated ages, we drew an isochrone of an appropriate age from the CMD table and performed a quadratic interpolation over the mass points within it to find the value for each parameter that corresponds to $M_{\mathrm{dyn}}$ at the given age.
This process was repeated for the upper and lower bounds on $M_{\mathrm{dyn}}$ to find the confidence bounds for each parameter at the given age. Our estimates for the stellar parameters using this method are presented in Table~\ref{tab:isochrones}. Figure~\ref{fig:hr_diagram} shows the isochrones for each age estimate plotted on a Hertzsprung-Russell diagram of $\log(L)$ vs $\log(T_\mathrm{eff})$, as well as the interpolated values for the primary and secondary components as large circles and triangles, respectively.

From Fig.~\ref{fig:hr_diagram}, it can be seen that, at the ages of 500\,kyr, 1\,Myr and 3\,Myr, the primary star is on the main sequence, with negligible variations in parameters between these ages. At the upper bound age of 10\,Myr, the primary is in the process of beginning to evolve beyond the main sequence. Conversely, it is clear from the HR diagram that the secondary component is predicted to be on the main sequence for all ages save the youngest, where it is still in its protostellar stage. Also on Fig.~\ref{fig:hr_diagram} is shown the location of the primary as calculated by \citet{Fairlamb15}, which appears to agree with all ages except the upper bound of 10~Myr. In Sect.~\ref{sec:age}, we generalise this process to all generated isochrones to attempt to constrain the age of the system.

\begin{table}
    \centering
    \caption{\texttt{PARSEC 1.2S / COLIBRI S\_37} models of each component of MWC\,166\,A. Isochrones were selected at five representative ages including reasonable lower and upper bounds, with stellar parameters corresponding to the dynamical masses of each component. The isochrones assume Solar metallicities and that the stars are coeval.}
    \begin{tabular}{l | S{c} | S{c} }
        \hline
        Parameter & MWC\,166\,Aa & MWC\,166\,Ab\\
        \hline
        $M_{\mathrm{dyn}}$ [$M_\odot$] & $12.19\pm2.18$ & $4.90\pm0.52$ \\
        \hline
        \hline
        {Age [yr]} & \multicolumn{2}{S{c}}{{$1.0\times10^5$}}\\
        \hline
        {$L$ [$L_\odot$]} & {$\left(1.20^{+0.53}_{-0.31}\right)\times10^{4}$} & {$\left(1.15^{+0.39}_{-0.24}\right)\times10^{2}$} \\
        {$R$ [$R_\odot$]} & {$4.43^{+0.11}_{-2.03}$} & {$13.66^{+1.80}_{-1.31}$} \\
        {$T_\mathrm{eff} $ [K]} & {$28~700^{+2400}_{-6600}$} & {$5~120^{+60}_{-40}$} \\
        {$\log(g)$} & {$4.23^{+0.02}_{-0.41}$} & {$2.85^{+0.04}_{-0.06}$}\\
        \hline
        \hline
        Age [yr] & \multicolumn{2}{S{c}}{$5.0\times10^5$}\\
        \hline
        $L$ [$L_\odot$] & $\left(1.06^{+0.67}_{-0.49}\right)\times10^{4}$ & $\left(3.77^{+8.01}_{-2.72}\right)\times10^{2}$ \\
        $R$ [$R_\odot$] & $4.21^{+0.44}_{-0.48}$ & $8.02^{+3.70}_{-1.17}$ \\
        $T_\mathrm{eff} $ [K] & $28~600^{+2200}_{-2500}$ & $9~000^{+7300}_{-2900}$ \\
        $\log(g)$ & $4.27\pm0.02$ & $3.32^{+0.58}_{-0.21}$\\
        \hline
        \hline
        Age [yr] & \multicolumn{2}{S{c}}{$1.0\times10^6$}\\
        \hline
        $L$ [$L_\odot$] & $\left(1.08^{+0.70}_{-0.50}\right)\times10^{4}$ & $\left(5.13^{+2.23}_{-1.72}\right)\times10^{2}$ \\
        $R$ [$R_\odot$] & $4.30^{+0.48}_{-0.50}$ & $2.46^{+0.44}_{-0.15}$ \\
        $T_\mathrm{eff} $ [K] & $28~400^{+2100}_{-2500}$ & $17~500^{+1100}_{-2900}$ \\
        $\log(g)$ & $4.25\pm0.02$ & $4.34^{+0.19}_{-0.01}$\\
        \hline
        \hline
        Age [yr] & \multicolumn{2}{S{c}}{$3.0\times10^6$}\\
        \hline
        $L$ [$L_\odot$] & $\left(1.16^{+0.80}_{-0.55}\right)\times10^{4}$ & $\left(5.13^{+2.23}_{-1.72}\right)\times10^{2}$ \\
        $R$ [$R_\odot$] & $4.59^{+0.58}_{-0.59}$ & $2.50\pm0.16$ \\
        $T_\mathrm{eff} $ [K] & $28~000^{+2100}_{-2500}$ & $17~400^{+1000}_{-1100}$ \\
        $\log(g)$ & $4.20\pm0.03$ & $4.33\pm0.01$\\
        \hline
        \hline
        Age [yr] & \multicolumn{2}{S{c}}{$1.0\times10^7$}\\
        \hline
        $L$ [$L_\odot$] & $\left(1.60^{+1.51}_{-0.86}\right)\times10^{4}$ & $\left(5.28^{+2.39}_{-1.79}\right)\times10^{2}$ \\
        $R$ [$R_\odot$] & $5.87^{+1.91}_{-1.22}$ & $2.61^{+0.20}_{-0.19}$ \\
        $T_\mathrm{eff} $ [K] & $26~800^{+700}_{-1900}$ & $17~100^{+1000}_{-1100}$ \\
        $\log(g)$ & $3.98^{+0.12}_{-0.17}$ & $4.29\pm0.02$\\
        \hline
    \end{tabular}
    \label{tab:isochrones}
\end{table}

\section{Results: Modelling the gas distribution \& kinematics in the He\,\textsc{i} and Br\,$\gamma$ line}
\label{sec:PMOIRED_results}

The circumprimary disc model described in Sect.~\ref{sec:linemodelling} produces an excellent fit to the data at all epochs.

\begin{figure*}
    \centering
    \includegraphics[width=\textwidth]{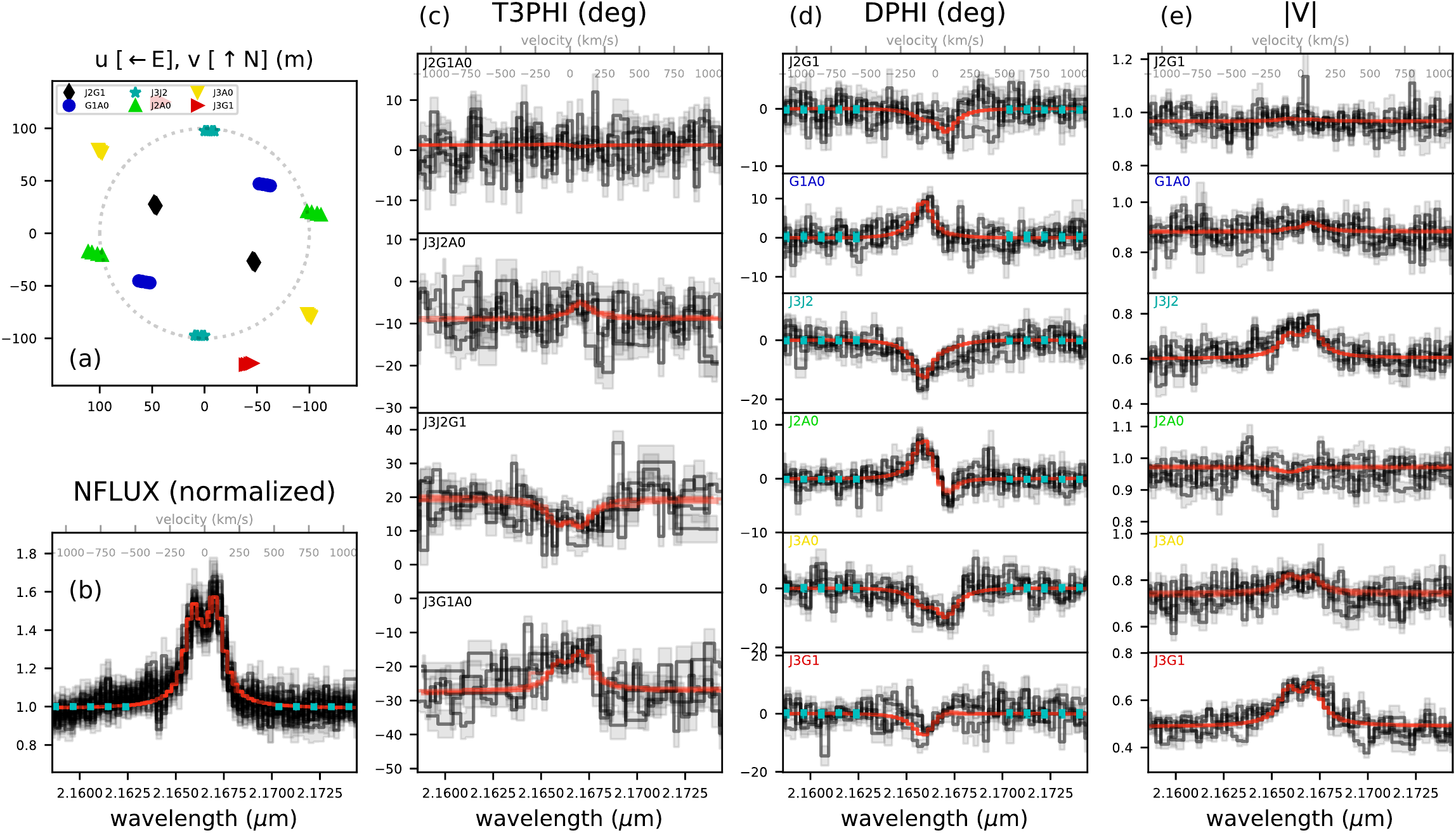}
    \vspace{1mm}
    \includegraphics[width=\textwidth]{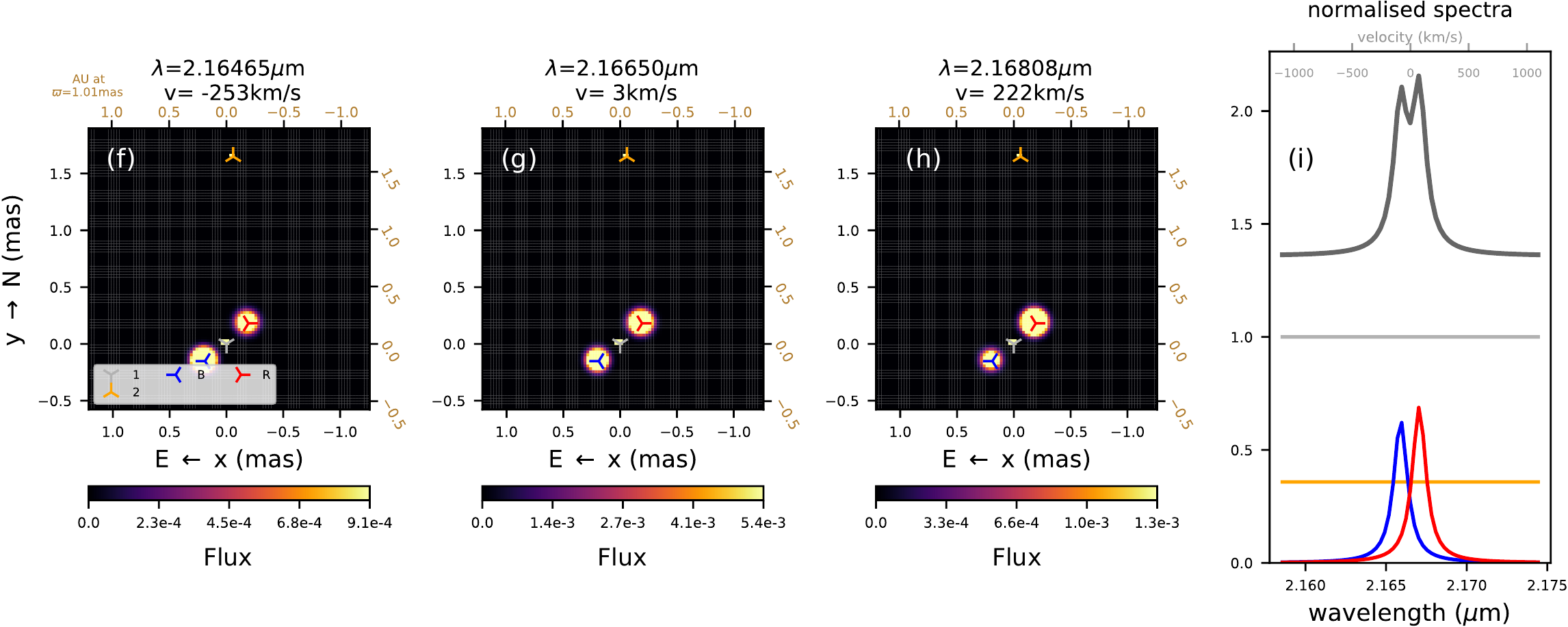}
    \caption{Results of the Br\,$\gamma$ line modelling, for the epoch 2017-03-14. \textbf{(a)}: The $(u,v)$-coverage for the observations associated with the epoch, coloured by baseline pair. \textbf{(b)}: Telluric-corrected normalised flux (labelled NFLUX, black lines), overplotted with flux in the best-fit model (red line). \textbf{(c)}, \textbf{(d)}, \textbf{(e)}: Data from each GRAVITY exposure (black lines), overplotted  with quantities computed from best-fit model (red lines). The observables are closure phase for each telescope triplet (T3PHI), differential phase for each baseline (DPHI), and visibility for each baseline (|V|), respectively. \textbf{(f)}, \textbf{(g)}, \textbf{(h)}: brightness distribution corresponding to the best fit, for three representative wavelengths. \textbf{(i)}: Synthetic line strengths and profiles for the two spectral components in the model (red and blue), in addition to the monochromatic continuum flux associated with the primary component (silver) and the secondary component (gold). The parameters corresponding to the fits can be found in Table~\ref{tab:PMOIRED_params}.}
    \label{fig:PMOIRED_model}
\end{figure*}

Figure~\ref{fig:PMOIRED_model} shows the results of the fit for the epoch 2017-03-14, where panel (a) shows the spectra, closure phases, differential phases, and visibilities measured with GRAVITY overplotted with prediction from the best-fit model. The three left-most plots in panel (b) show synthetic images computed from the best-fit model for three representative wavelengths. The central panel shows the brightness distribution at $\lambda_0$, while on the left- and right-hand sides of this, the wavelength is red-shifted and blue-shifted respectively, by a factor of $\lambda_0\pm3\Delta\lambda$. The right-most panel of the figure shows the flux contribution of each synthetic model component.

\begin{table}
    \caption{Model parameters corresponding to the best-fit \texttt{PMOIRED} circumprimary disc line models, for both spectral lines of interest, for each GRAVITY epoch (Sect.~\ref{sec:PMOIRED_results}).}
    \centering
    \begin{adjustbox}{angle=90}
    \begin{tabular}{|l || c | c || c | c || c | c || c | c|}
    \hline
    Epoch & \multicolumn{2}{c||}{2017-03-14} & \multicolumn{2}{c||}{2017-04-27} & \multicolumn{2}{c||}{2018-01-11} & \multicolumn{2}{c|}{2018-02-06} \\
    \hline
    Spec. Line & He\,\textsc{i} & Br\,$\gamma$ & He\,\textsc{i} & Br\,$\gamma$ & He\,\textsc{i} & Br\,$\gamma$ & He\,\textsc{i} & Br\,$\gamma$ \\
    \hline
    \hline
    $\rho$ [mas] & $1.67\pm0.01$ & $1.65\pm0.01$ & $1.98\pm0.01$ & $1.98\pm0.01$ & $0.86\pm0.02$ & $0.90\pm0.02$ & $1.59\pm0.01$ & $1.56\pm0.01$ \\
    $\theta$ [$^\circ$] & $357.4\pm0.3$ & $357.8\pm0.3$ & $42.3\pm0.3$ & $42.0\pm0.2$ & $238.1\pm1.1$ & $238.9\pm1.0$ & $307.8\pm0.2$ & $307.3\pm0.2$ \\
    $f_2$ & $0.414\pm0.004$ & $0.359\pm0.003$ & $0.457\pm0.004$ & $0.402\pm0.002$ & $0.318\pm0.025$ & $0.240\pm0.016$ & $0.440\pm0.007$ & $0.399\pm0.006$ \\
    \hline
    \hline
    $x_B$ [mas] & $0.212\pm0.017$ & $0.192\pm0.015$ & $0.242\pm0.034$ & $0.255\pm0.031$ & $0.114\pm0.016$ & $0.223\pm0.022$ & $0.175\pm0.013$ & $0.161\pm0.013$ \\
    $y_B$ [mas] & $-0.118\pm0.017$ & $-0.184\pm0.012$ & $-0.226\pm0.028$ & $-0.118\pm0.024$ & $-0.091\pm0.023$ & $-0.076\pm0.036$ & $-0.052\pm0.011$ & $-0.123\pm0.011$\\
    $\sigma_B$ [mas] & $0.121$ & $0.133$ & $0.166$ & $0.141$ & $0.073$ & $0.118$ & $0.091$ & $0.102$ \\
    $f_B$ & $0.712\pm0.015$ & $0.633\pm0.015$ & $0.668\pm0.026$ & $0.497\pm0.023$ & $0.533\pm0.016$ & $0.289\pm0.020$ & $0.945\pm0.016$ & $0.643\pm0.015$\\
    \hline
    \hline
    $x_R$ [mas] & $-0.155\pm0.011$ & $-0.190\pm0.013$ & $-0.168\pm0.021$ & $-0.208\pm0.021$ & $-0.048\pm0.017$ & $0.040\pm0.020$ & $-0.130\pm0.015$ & $-0.053\pm0.020$\\
    $y_R$ [mas] & $0.168\pm0.009$ & $0.135\pm0.011$ & $0.188\pm0.017$ & $0.202\pm0.017$ & $0.184\pm0.026$ & $0.244\pm0.032$ & $0.241\pm0.014$ & $0.243\pm0.019$\\
    $\sigma_R$ [mas] & $0.114$ & $0.117$ & $0.126$ & $0.145$ & $0.095$ & $0.124$ & $0.137$ & $0.124$\\
    $f_R$ & $1.074\pm0.014$ & $0.688\pm0.013$ & $1.009\pm0.021$ & $0.664\pm0.017$ & $0.425\pm0.014$ & $0.309\pm0.016$ & $0.603\pm0.014$ & $0.326\pm0.012$\\
    \hline
    \hline
    $F_\mathrm{L}$ & $1.045\pm0.016$ & $1.046\pm0.022$ & $1.231\pm0.031$ & $1.340\pm0.041$ & $1.150\pm0.026$ & $1.386\pm0.048$ & $1.051\pm0.018$ & $1.330\pm0.030$\\
    $\lambda_0$ [${\upmu}$m] & $2.0590$ & $2.1665$ & $2.0593$ & $2.1667$ & $2.0593$ & $2.1667$ & $2.0592$ & $2.1667$\\
    $\Delta\lambda$ [${\upmu}$m] & $0.617\pm0.005$ & $0.577\pm0.006$ & $0.568\pm0.009$ & $0.558\pm0.011$ & $0.578\pm0.008$ & $0.422\pm0.016$ & $0.606\pm0.006$ & $0.576\pm0.011$\\
    \hline
    \hline
    $\chi^2$ & 2.55 & 2.37 & 4.18 & 3.07 & 2.22 & 1.95 & 3.53 & 3.04 \\
    \hline
    \end{tabular}
    \end{adjustbox}
    \label{tab:PMOIRED_params}
\end{table}

The line modelling results for all epochs (analogously to Fig. \ref{fig:PMOIRED_model}) are shown in the Appendix, and the findings are summarised in Table \ref{tab:PMOIRED_params}, for both spectral lines of interest.\\

The results show substantial evidence of line emission being localised around the primary star, consistent throughout all four GRAVITY epochs. The red-shifted component (R) is displaced consistently towards the north-west of the primary star, while the blue-shifted component (B) is located south-east of the primary. The displacement of the two components from the star is generally {(\textasciitilde0.25\,mas)}. The average displacement for He\,\textsc{i} was found to be $(10.5\pm0.4)\,R_1$, and for the Br\,$\gamma$ this was $(11.5\pm0.4)\,R_1$. The derived values for each epoch can be seen in column (5) of Table~\ref{tab:RBflux_andRho}.
The relative intensities of the $B$ and $R$ components also seem to be variable by epoch and by spectral line. From columns (3), (4) of Table~\ref{tab:RBflux_andRho}, it can be seen that, for the He\,\textsc{i} line, the red wing contributes 60\% of the flux for the first two epochs, a similar level of intensity as the blue wing during the latter two epochs. On the other hand, the Br\,$\gamma$ line flux tends to be more equally distributed between the two components, apart from in the epoch 2018-02-06, where the blue component is substantially stronger.
The emission of the red- and blue-shifted line wing are displaced along an average position angle of $134.0\pm1.1^\circ$, indicating that this is the sky-projected position angle of the major-axis of the rotating disc. Accordingly, the rotation axis of the disks, and likely also the primary star, is oriented along position angle $44.0\pm1.1^\circ$ on sky.



\begin{table*}
    \caption{Columns (3) and (4): relative strengths of the red and blue wings of the respective spectral line. Uncertainties are 0.01 for all values in these columns. Columns (5) and (6): radius of line emission centre, averaged from the red and blue wings, given respectively in units of milli-arcseconds, and the primary star's radius at 3 Myr (see Table~\ref{tab:isochrones}). Column (7) shows the estimated rotational axis of the primary star.}
    \centering
    \begin{tabular}{c | c || c | c | c | c | c}
        \hline
        Epoch & Line & $f_R/f_{R+B}$ & $f_B/f_{R+B}$ & $\bar{\rho}_\mathrm{line}$ [mas] & $\bar{\rho}_\mathrm{line} \, [R_1]$ & $\theta_\mathrm{rot}\,[^\circ]$\\
        (1) & (2) & (3) & (4) & (5) & (6) & (7)\\
        \hline
        2017-03-14 & He\,\textsc{i} & 0.60 & 0.40 & $0.24\pm0.02$ & $11.2\pm0.9$ & $37.9\pm2.4$ \\   
                   & Br\,$\gamma$ & 0.52& 0.48 & $0.25\pm0.01$ & $11.6\pm0.5$ & $39.9\pm2.1$\\
        \hline
        2017-04-27 & He\,\textsc{i} &0.60 & 0.40 & $0.29\pm0.03$ & $13.5\pm1.4$ & $45.0\pm4.0$\\  
                   & Br\,$\gamma$ & 0.57 & 0.43 & $0.29\pm0.03$ & $13.5\pm1.4$ & $34.7\pm3.3$\\
        \hline
        2018-01-11 & He\,\textsc{i} & 0.44 & 0.56 & $0.17\pm0.02$ & $7.9\pm0.9$ & $59.0\pm5.0$\\
                   & Br\,$\gamma$ & 0.52 & 0.48 & $0.24\pm0.03$ & $11.2\pm1.4$ & $60.0\pm5.0$\\
        \hline
        2018-02-06 & He\,\textsc{i} & 0.39 & 0.61 & $0.23\pm0.01$ & $10.7\pm0.5$ & $43.9\pm2.5$\\
                   & Br\,$\gamma$ & 0.34 & 0.66 & $0.23\pm0.02$ & $10.7\pm0.9$ & $59.7\pm3.2$\\
        \hline
    \end{tabular}
    \label{tab:RBflux_andRho}
\end{table*}

\section{Discussion}
\label{sec:discussion}

\subsection{Evidence for circumbinary dust}
\label{sec:circumstellar}

Our interferometric observations in the continuum allow us to quantify the excess emission through our visibility fit (Sect.~\ref{sec:extended_geometry}), where we find that {\textasciitilde2\%} and {\textasciitilde5\%} of the excess emission are associated with extended flux in the $H$-band and $K$-band, respectively. The $H$-band observations also show `spikes' in the extended flux contribution in the epochs 2019-12-24 and 2020-11-19, with $f_\mathrm{ext}/f_\mathrm{tot}$ contributing up to $15\%$ of the total flux for these epochs. The geometry of the extended dust is still poorly constrained, but our modelling shows the emission originates from scales at least twice larger than the binary separation vector (for a ring model), but likely even larger (for a Gaussian or background flux model; Table~\ref{tab:MIRCXmodels} and Sect.~\ref{sec:extended_geometry}). This is comparable to the dynamical truncation radius predicted by \citet{arty+lubow94} for circular binaries ({\textasciitilde$1.7a$}).

The expected Silicate dust sublimation radius ($R_\mathrm{s}$) can be estimated with 
\begin{equation}
    R_\mathrm{s} =  1.1 \sqrt{Q_\mathrm{R}}\;\Bigg(\frac{L}{1000\, L_\odot}\Bigg)^{1/2} \Bigg(\frac{T_\mathrm{s}}{1500\, \mathrm{K}}\Bigg)^{-2}\; \mathrm{au},
\label{eqn:dustsublim}
\end{equation}
where $L$ is the bolometric luminosity of the star(s) irradiating the disk and $T_s$ is the dust sublimation temperature \citep{monnier_millan-gabet02}. $Q_\mathrm{R} \equiv Q_\mathrm{abs}(T_*)/Q_\mathrm{abs}(T_\mathrm{s})$ is the ratio of dust absorption efficiencies for radiation for the incident and re-emitted field, which we fix to $Q_\mathrm{R} = 1$ in order to estimate the rim  location for large $\upmu$m-sized dust grains. The above equation has several simplifications compared to the physical system, chief among which is the assumption of a single star rather than a binary. While dust sublimation radii have been calculated for binary systems \citep[e.g.][]{nagel10}, dynamical interactions cause the inner dust rim to be at a larger radius than $R_s$. A detailed calculation is therefore outside the scope of this work, but equation \ref{eqn:dustsublim} can still give an order-of-magnitude estimate for the minimum dust inner rim radius of the circumbinary disc, assuming a luminosity equivalent to that of the two components combined. Substituting the sum of the luminosity values from the central age estimate of 1\,Myr in Table~\,\ref{tab:isochrones} gives $R_\mathrm{s} = 2.6\,\mathrm{au}, 4.6\,\mathrm{au}$ and $7.1\,\mathrm{au}$ for dust sublimation temperatures of 2000\,K, 1500\,K and 1200\,K, respectively. This suggests that individual dusty circumstellar discs are likely not present in the MWC\,166\,A system. Due to the calculated semi-major axis of the orbit being comparable to the smallest of the above estimates ($a_1=2.61\,\mathrm{au}$ at $d=990\,\mathrm{pc}$), it is expected that substantial individual dusty discs around each of the components are not capable of surviving for extended periods of time due to dynamical interactions \citep[e.g.][]{mathieu94}. Accordingly, we associate the extended continuum emission with a circumbinary disc component instead of circumstellar disc component.

\subsection{Evidence for variable extinction or circumstellar material}
\label{sec:variability}

The relative flux contribution of the two point sources in our model was found to be variable, especially in the $H$-band continuum. The flux associated with the secondary appears to increase around phase \textasciitilde0.9, and was found to be as bright as the primary at one epoch (2019-12-24) and even brighter than the primary at two epochs (2019-12-16 and 2020-12-28). These latter two epochs were both taken on the compact AT configuration, and returned visibilities very close to unity -- as well as closure phases close to zero -- at all probed baselines (See Figs. \ref{fig:vis_all} and \ref{fig:cp_allepochs_sf}). This, and the more limited $(u,v)$-coverage caused by the shorter baselines, can result in ambiguities in the model \citep{anthonioz15}. To rule this out, we repeated the modelling for these two epochs while restricting the secondary to a maximum brightness of 100\% of the primary. We then repeated the astrometric fit, and found it to be incompatible with any physical orbit when considered with the other points. This means that the secondary's increase in relative $H$-band brightness to outshine the primary at the epochs 2019-12-16 and 2020-12-28 is the only solution which agrees with the Keplerian orbit of the objects.

Since these two points are at very similar orbital phases (0.88 and 0.91 respectively), we argue that this variability could be a result of the dynamical interaction of the secondary with the circumbinary disc, possibly through variable extinction, where the line-of-sight extinction changes towards one of the stars due to rearrangements in the circumbinary disc. In addition, our point-source flux estimate ($f_2/f_1$) might contain emission contributions from circumstellar gas or dust, in particular as the binary separation is still comparable to our interferometric beam size. Therefore, the observed $f_2/f_1$ brightness increase might correspond to an increase in excess emission near the location of the secondary as it approaches periastron, either through an accretion burst onto the secondary or from a hot spot at the inner edge of the circumbinary disc caused by the additional heating from the secondary. Indeed, SPH simulations of young, eccentric close binaries with circumbinary discs such as those by \citet[][for HD\,104237]{Dunhill15} and \citet[][for DQ\,Tau]{muzerolle19}, have suggested that dynamical interactions can cause a differential rate of accretion depending on the orbital phase, with the highest accretion peak occurring during the $10-20\%$ of the orbit preceding periastron. The observed brightening, around orbital phase \textasciitilde0.9, could be consistent with this prediction.

\subsection{Nature of the line-emitting region}

As shown in Sect.~\ref{sec:PMOIRED_results}, the geometry in the $K$-band emission lines is substantially different from what we see in the continuum. The best fit to the circumstellar environment in the He\,\textsc{i} and Br\,$\gamma$ lines indicates a strongly-emitting Keplerian disc around the primary component. In this section, we present three possible interpretations for the physical origin of the line emission, either emission from the accretion region, an ionised gas accretion disc channelling circumstellar material to the star, or a decretion disc tracing material from the star being lost through stellar winds.

\subsubsection{Accretion onto the primary}
\label{sec:accretion}

Brackett\,$\gamma$ emission is a common marker of magnetospheric accretion in YSOs and is especially useful for accretion rate determination, since the luminosity of the line ($L_\mathrm{Br\gamma}$) has been shown to be directly related to the accretion luminosity $L_\mathrm{acc}$ over a wide mass range from brown dwarfs to Herbig~Ae stars \citep[e.g.][]{muzerolle98, calvet04}. This relation follows a power law which is strongly dependent on stellar mass \citep{fairlamb17}.

For Herbig~Ae objects, \citet{donehew11} derived the following relation:
\begin{equation}
    \log(L_\mathrm{acc}/L_\odot) = (0.9\pm0.2)\log(L_\mathrm{Br\gamma}/L_\odot)+(3.3\pm0.7),
    \label{eqn:lacc}
\end{equation}
with similar values obtained by \citet{mendigutia11}. From the above relation, the accretion rate $\dot{M}$ can be derived:
\begin{equation}
    L_\mathrm{acc} = \frac{G M_* \dot{M}}{R_*},
    \label{eqn:lacc_to_mdot}
\end{equation}
where $M_*$ and $R_*$ are respectively the mass and radius of the star. Using the Br\,$\gamma$ luminosity for MWC\,166\,A calculated by \citet{donehew11} of $L_\mathrm{Br\gamma}~=~(11\pm5)\times10^{-3}\,L_\odot$, we can estimate the accretion luminosity and hence the mass accretion rate. This results in $L_\mathrm{acc}\sim35\pm65 L_\odot$ and subsequently ${\dot{M}\sim(3.9\pm7.4)\times10^{-7}\,M_\odot\,\mathrm{yr^{-1}}}$, which is in line with the typical mass accretion rate of $2\times10^{-7}\,M_\odot\,\mathrm{yr^{-1}}$ found by \citet{mendigutia11}.

However, it must be noted that, with a mass of 12 $M_\odot$, MWC\,166\,Aa is a Be star, leaving it outside of the regime where the $L_\mathrm{Br\gamma}$ -- $L_\mathrm{acc}$ relation has been calibrated. Indeed, Eq.~\ref{eqn:lacc} provides a systematic overestimate of $L_\mathrm{acc}$ for Be stars, according to \citet{donehew11}. Furthermore, the large uncertainties inherent to the estimate of $L_\mathrm{Br\gamma}$ by \citet{donehew11} result in very large errors on the propagated quantities. As a result, the above value for $\dot{M}$ should be taken to be an order-of-magnitude estimate.

The strong Br\,$\gamma$ signal is consistent with accretion, as is the youth of the system. However, the geometry of the line emission does not appear to be consistent with direct stellar accretion. The radius of the emission originates from a region further from the stellar surface than the $\sim5R_\star$ which would be expected from accretion in Herbig Ae/Be stars \citep{bouvier20}. Furthermore, the geometry of the line emission is also stable over all epochs, with the red and blue lobes consistently located north-west and south-east of MWC\,166\,Aa, and with velocities consistent with Keplerian rotation, which is not expected of material being funnelled onto a stellar surface \citep{bouvier07}. As a result, the line emission does not appear to be tracing direct stellar accretion, but rather a process which is more spatially extended.


\subsubsection{Inner gas accretion disc}
\label{sec:accretion_disc}

The line emission might trace ionised gas in the inner region of the circumprimary gas disc. Based on hydrodynamic simulations, we expect that a stable circumprimary gas disc can exist out to one third of the binary separation \citep{arty+lubow94}. This disc could accommodate the mass transport from the large-scale mass reservoir seen in far-infrared excess emission, over a circumbinary disc to the star. For MWC\,166\,A, the upper limit on the radius is {\textasciitilde0.9\,au} or \textasciitilde$15\,R_1$, which is in agreement with the observed location of the Br\,$\gamma$ emission (\textasciitilde$11.5\,R_1$). These discs are often found around Herbig Ae/Be systems, which would be consistent with the age of the system (see Sect.~\ref{sec:age}). The emission would then be tracing ionised gas in Keplerian rotation around the star, rather than direct accretion onto the star.
\citet{Kraus12b} found an example of such a disc around another young B-type close binary system, V921 Sco, using similar techniques. The Br\,$\gamma$ line profiles were found to be similarly strong and narrow to those presented in this study for MWC\,166\,Aa. They also were similarly variable in intensity and wing strength, while still being consistent with a Keplerian disc.

However, the lack of evidence of direct accretion in our observations means that this scenario cannot be confirmed. Further observations at shorter wavelengths may show evidence of accretion, for example by showing emission closer to the stellar surface.


\subsubsection{Be decretion disc}
\label{sec:decretion_disc}
As a third scenario, we consider that the observed circumprimary emission might be associated with a decretion disc, as observed around classical Be stars. Classical Be stars are defined as non-supergiant B stars which have shown Balmer line emission at some point in time \citep{collins87}. Generally very fast rotators, they typically host gaseous circumstellar discs characterised by a viscous decretion model \citep[e.g.][]{lee91,carciofi11}, with material ejected by radiation pressure from the star forming a Keplerian disc. This disc is sustained by periodic outbursts from the star \citep{grundstrom11}, which leads to variability on a range of timescales from days to years \citep{labadieBartz17}.


Variability in the emission line profile is also common in decretion discs, with violet-to-red (V/R) variations meaning the red-shifted and blue-shifted wings of the line are commonly found at different relative strengths at different epochs \citep{rivinius13} -- Fig.~\ref{fig:flux} shows that such variations are indeed present in the spectrum of MWC\,166\,A, and the results in columns (3) and (4) of Table~\ref{tab:RBflux_andRho} quantify this. In the simplest of Keplerian models, the flux intensities of each wing should be equal. The observed deviation from this profile is due to the flux intensity of the disc being azimuthally asymmetrical \citep{porter03}, which is most likely due to temperature or gas density enhancements in specific regions of the disc. Our observations, comprising only four epochs, are insufficient to characterise the observed variations in V/R ratio as being periodic or quasi-periodic, but previous studies \citep{hanuschik95} indicate that, in general, these variations are cyclic in the long-term. \citet{hummel97} modelled a one-armed density wave precessing around the star as an explanation for these variations.

The rapidly changing V/R ratio is characteristic of Classical Be stars observed at intermediate inclinations \citep{catanzaro13}. Such variations are generally not found in Herbig Be objects. Furthermore, the line-emitting region around the primary is rather extended, which is more consistent with decretion than accretion. We would assume Br\,$\gamma$ emission from magnetospheric accretion to originate no more than $5 R_\star$ from the star \citep{bouvier20}. However, we find that at all epochs, it is located well beyond the stellar surface. We estimated the radius of the emission by averaging the angular displacement in the blue and red wings of each line, with the results shown in column (5) of Table~\ref{tab:RBflux_andRho}. These values do agree with each other in almost all circumstances, despite some slight variation between the two lines and epochs, with a general separation from the primary of {\textasciitilde0.25\,mas}, or {\textasciitilde$11{R_1}$}. Comparing this value to other spectro-interferometric Be studies shows that it is consistent with that of a Classical Be disc, with the VLTI/AMBER survey by \citet{cochetti19} showing Br\,$\gamma$ radii ranging between {$3-13\,R_\star$}. The lack of dust in Be discs also correlates with our findings from the continuum fit, where we found no evidence for circumprimary dust.

In light of the above points, the decretion disc model {cannot be excluded for the circumprimary emission.} Previous photometric studies of MWC\,166\,A found evidence for periodic dissipation and regeneration of the circumstellar envelope around the primary star over a period of months \citep{Pogodin06}, which is consistent with models and observations of Be stars and adds weight to this interpretation of the line emission.

This would be in contrast to previous characterisations of MWC\,166\,Aa as a Herbig Be star \citep{Fairlamb15}, and is also supported by our isochrone interpolation: from Fig.~\ref{fig:hr_diagram}, it can be seen that the large mass of the primary means that it is evolved beyond the protostellar stage at all age estimates. This is further supported by the relatively small near-infrared excess in the spectral energy distribution of MWC\,166\,A, suggesting that the circumstellar environment has been at least partially cleared of material. The observed far-infrared excess of the system at \textasciitilde$100\,\upmu\mathrm{m}$, however, is more supportive of the accretion disc scenario outlined in Sect.~\ref{sec:accretion_disc}.

As a result, we can characterise the $K$-band line emission as originating in a Keplerian or quasi-Keplerian circumprimary gas disc, as opposed to direct stellar accretion. Whether this is an accretion disc or a decretion disc is to be confirmed by future observations.


\subsection{Constraining the age of the system}
\label{sec:age}
As Table~\ref{tab:isochrones} and Fig.~\ref{fig:hr_diagram} show, the primary star has already reached the main-sequence for all plausible age estimates, while the secondary is still in its pre-main-sequence stage at the youngest age estimates. By comparing the luminosities of the two stars at different ages to the flux ratios found from our interferometric observations, we can place some constraints on the age of the system.

\begin{figure}
    \centering
    \includegraphics[width=\columnwidth]{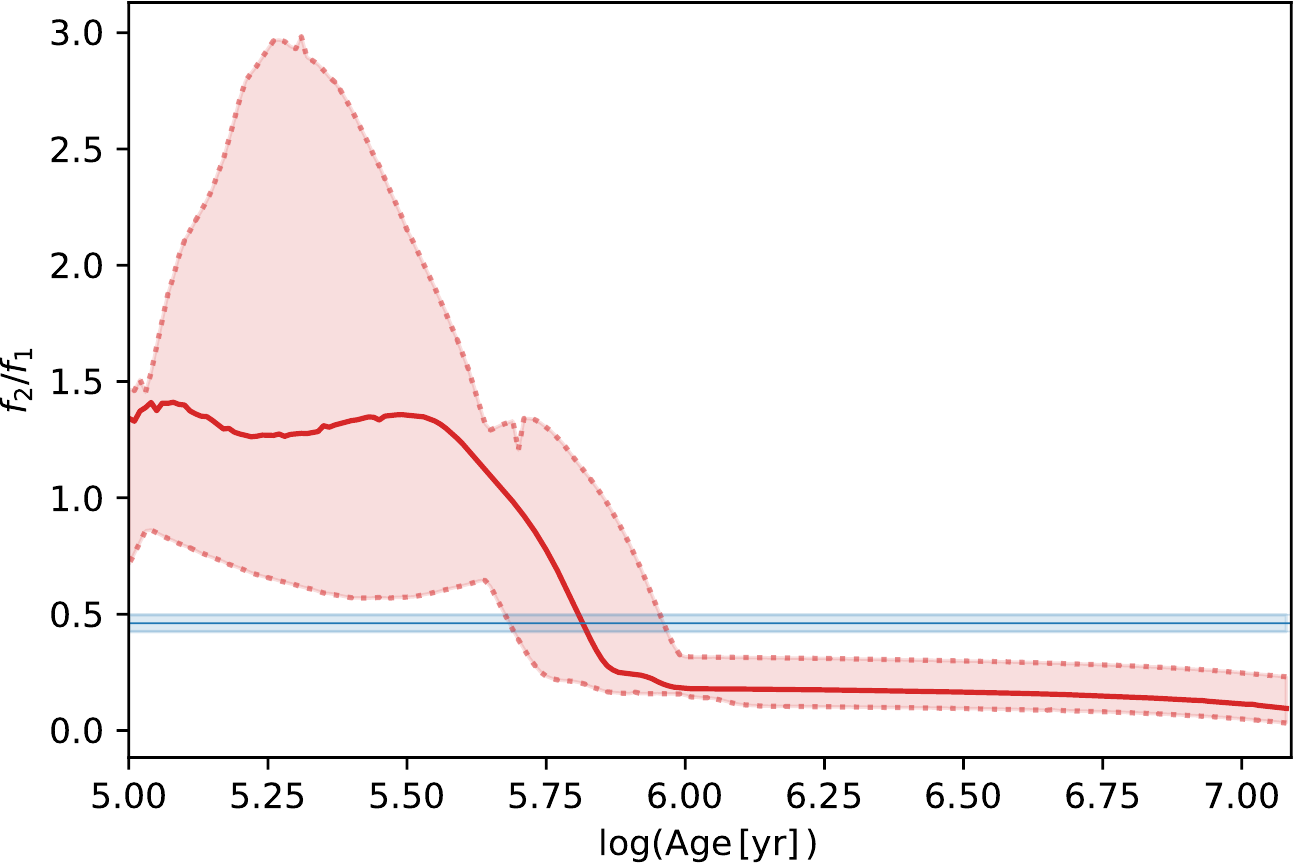}
    \includegraphics[width=\columnwidth]{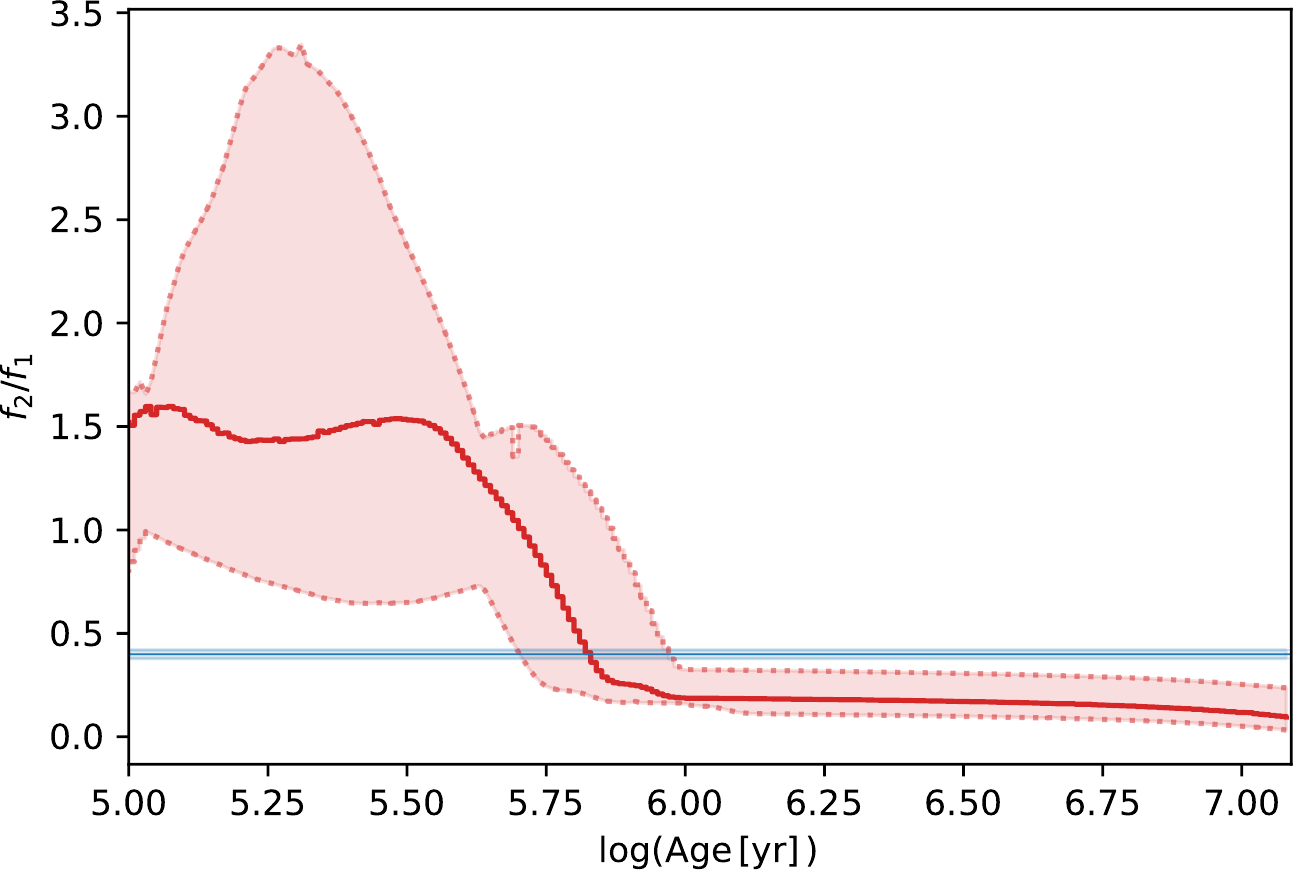}
    \caption{Flux ratio $f_2/f_1$ plotted against $\log(\mathrm{Age})$, for all isochrones (red lines), compared to interferometric flux ratios at the same wavelength (blue lines). The upper panel shows the flux ratio at $1.65\,\upmu\mathrm{m}$, while the lower panel shows the flux ratio at $2.2\,\upmu\mathrm{m}$ -- the central wavelengths for the $H$- and $K$-bands respectively. The lighter-shaded areas illustrate the uncertainty on each quantity.}
    \label{fig:fluxratiovsage}
\end{figure}

We first performed the process described in Sect.~\ref{sec:stellarproperties} for every age isochrone in our CMD table between 100\,kyr and 10\,Myr. We then used the derived parameters to generate spectral energy distributions for both stars, using model atmospheres from \citet{kurucz93}. We then calculated values of $f_2/f_1$ at the central wavelengths of both the $H$- and $K$-bands, ${1.65\,\upmu{\mathrm{m}}}$ and ${2.2\,\upmu{\mathrm{m}}}$ respectively, to enable a comparison with the corresponding values from our continuum interferometry. The resulting flux ratios are plotted as the red lines on Fig.~\ref{fig:fluxratiovsage}, with the $H$-band ratios in the upper panel and the $K$-band ratios in the lower panel.

The $H$-band flux ratio measurements show significant scatter between epochs, which might be due to scattered light contributions that are are important towards shorter wavelengths. As described in Sect.~\ref{sec:variability}, PIONIER will likely be more strongly affected by these contributions, while MIRC-X and the long CHARA baselines should be able to resolve out extended flux and separate the components more reliably. Therefore, we adopt the MIRC-X value of $f_2/f_1=0.484\pm0.014$ as our most reliable H-band flux ratio measurements, and mark this value with the blue line in the lower panel of Fig.~\ref{fig:fluxratiovsage}.

The $K$-band continuum GRAVITY data shows much less variability, and therefore we chose to take the weighted average of the GRAVITY flux ratios is plotted on the lower panel of Fig.~\ref{fig:fluxratiovsage}, corresponding to an average flux ratio of {${f_2/f_1 = 0.399\pm0.021}$}.

Comparing the measured $H$ and $K$-band flux ratio with the model values, we can exclude ages for the system of {$\lesssim500$\,kyr}. In the $K$-band, this is well beyond the  3$\sigma$ significance level. For the $H$-band, this is only 2$\sigma$, but the range of ages <500\,kyr for which this is the case is very small, and only due to our relatively large mass uncertainties -- which are themselves large because of the uncertainty on the system's distance.

At such young ages, the near-infrared brightness of the secondary would exceed the primary, due to the low temperature of the secondary component, which is not supported by our interferometric observations. The upper age of the system is less well constrained, as the stars evolve very slowly once they reach the main-sequence (see Fig.~\ref{fig:hr_diagram}). This can be seen in Fig.~\ref{fig:fluxratiovsage}, where the models predict consistently $f_2/f_1\sim0.2$ for ages beyond 1\,Myr. Our average continuum $K$-band flux ratio is larger than this, suggesting that the secondary component of the system is still in its pre-main-sequence stage. From our combined $H$ and $K$-band constraints, we estimate the system age to {$(7\pm2)\times10^5$\,yr}.
This young age is also consistent with the presence of circumstellar material (see Sect.~\ref{sec:circumstellar}). Such an age is also consistent with either the accretion or decretion disc models described in Sects.~\ref{sec:accretion_disc} and \ref{sec:decretion_disc}.

\section{Conclusions}
\label{sec:conclusions}

In this study we have presented GRAVITY, PIONIER and MIRC-X observations of MWC\,166\,A that resolve the system in the near-infrared $H$-band and $K$-band with milliarcsecond resolution. We derived the astrometry of the system at 13 epochs and calculated a first fully three-dimensional orbital solution for the system. This orbit differs substantially from the RV-only orbits of \citet{Corporon99} and \citet{Pogodin06}, having a period twice as long. We subsequently constrained the dynamical system mass ($17.1\pm2.7\,M_\odot$ for the photometric distance of {$990\pm50$\,pc} found by \citealt{kaltcheva00}) and the distance of the system, with our results excluding previous parallax measurements of the distance where $d<500$~pc. Furthermore, we have calculated, for the first time, the individual masses of the primary and the secondary components of the system, which we found to be ${M_1=(12.19\pm2.18)\,M_\odot}$ and ${M_2=(4.90\pm0.52)\,M_\odot}$ respectively. We also found estimates for the other fundamental stellar parameters based on quadratic isochrone interpolation.

Furthermore, we see evidence for circumstellar emission, both in the dust continuum and in the He\,\textsc{i} and Br\,$\gamma$ spectral lines, although they have different geometries. The geometry of the extended emission in the continuum is not well constrained, with the best fit corresponding to an overresolved background halo. The variability of the continuum emission between epochs may be an indication of physical variability in the quantity of circumstellar dust, or might indicate that the geometry is more complex than assumed in our model. Characterising it in more detail will require additional interferometric observations, ideally at mid-infrared wavelengths.

On the other hand, the geometry of the He\,\textsc{i} and Br\,$\gamma$ line emission is well constrained by our observations. Our models show emission in these lines to be localised around the primary, where the red-shifted and blue-shifted wings are spatially displaced, consistent with gas in a circumprimary disc.  The large spatial extend of the line-emitting regions (11.5\,$R_1$ for Br$\gamma$, 10.5\,$R_1$ for He\,\textsc{i}) and stable position angle orientation are inconsistent with an origin in magnetospheric accretion or boundary-layer accretion, but support the hypothesis that the line emission is tracing an ionised gas disc. This gas disc might either be fed by mass infall from outside the binary, or represent a decretion disc forming through mass-loss from the primary.

Finally, we constrain the age of the system to $(7\pm2)\times10^5$\,yr, based on the measured flux ratio of the components. We find that the primary is a main-sequence Be star, while the secondary is a Herbig Be object still in the process of gravitational contraction onto the main sequence.

\begin{acknowledgements}
    We acknowledge support from an STFC studentship (No.\ 109106G), an STFC Consolidated Grant (ST/V000721/1), and European Research Council Starting Grant "ImagePlanetFormDiscs" (Grant Agreement No.\ 639889) and Consolidated Grant "GAIA-BIFROST" (Grant Agreement No.\ 101003096). Travel support was provided by STFC PATT grant ST/S005293/1. MIRC-X received funding from the European Research Council (ERC) under the European Union's Horizon 2020 research and innovation programme (Grant No. 639889). JDM acknowledges funding for the development of MIRC-X (NASA-XRP NNX16AD43G, NSF-AST 1909165) and MYSTIC (NSF-ATI 1506540, NSF-AST 1909165).
    
    This research has made use of the VizieR catalogue access tool, CDS, Strasbourg, France. The original description of the VizieR service was published in \cite{Vizier}. This research has made use of the Jean-Marie Mariotti Center JSDC catalogue \footnote{Available at: \url{http://www.jmmc.fr/catalogue_jsdc.htm}}.  This research has made use of the Jean-Marie Mariotti Center OiDB service available at \url{http://oidb.jmmc.fr}.
    
    Analysis in this work was based on observations collected at the European Organisation for Astronomical Research in the Southern Hemisphere under ESO programme(s) 098.C-0910(A) (GRAVITY); 190.C-0963(A), 190.C-0963(B), 102.C-0701(B), 104.C-0737(A), 104.C-0737(B), 104.C-0737(C), 106.21JU.001, 106.21JU.002, 106.21JU.003 (PIONIER).

    This research has made use of the \texttt{PIONIER data reduction package} of the Jean-Marie Mariotti Center\footnote{Available at \url{http://www.jmmc.fr/pionier}}. This work has made use of data from the European Space Agency (ESA) mission {\it Gaia} (\url{https://www.cosmos.esa.int/gaia}), processed by the {\it Gaia} Data Processing and Analysis Consortium (DPAC, \url{https://www.cosmos.esa.int/web/gaia/dpac/consortium}). Funding for the DPAC has been provided by national institutions, in particular the institutions participating in the {\it Gaia} Multilateral Agreement. We thank the referee for their insightful comments
\end{acknowledgements}

\bibliographystyle{aa}
\bibliography{references}

\onecolumn
\appendix
\label{sec:appendix}
\section{\texttt{PMOIRED} line models}

Below we show the full set of He\,\textsc{i} and Br\,$\gamma$ line models for the four GRAVITY epochs, discussed in Sects.~\ref{sec:linemodelling} and \ref{sec:PMOIRED_results}. The figures comprise nine panels, laid out equivalently to those in Fig.~\ref{fig:PMOIRED_model}. 



\begin{figure}[H]
\centering
\includegraphics[width=\textwidth]{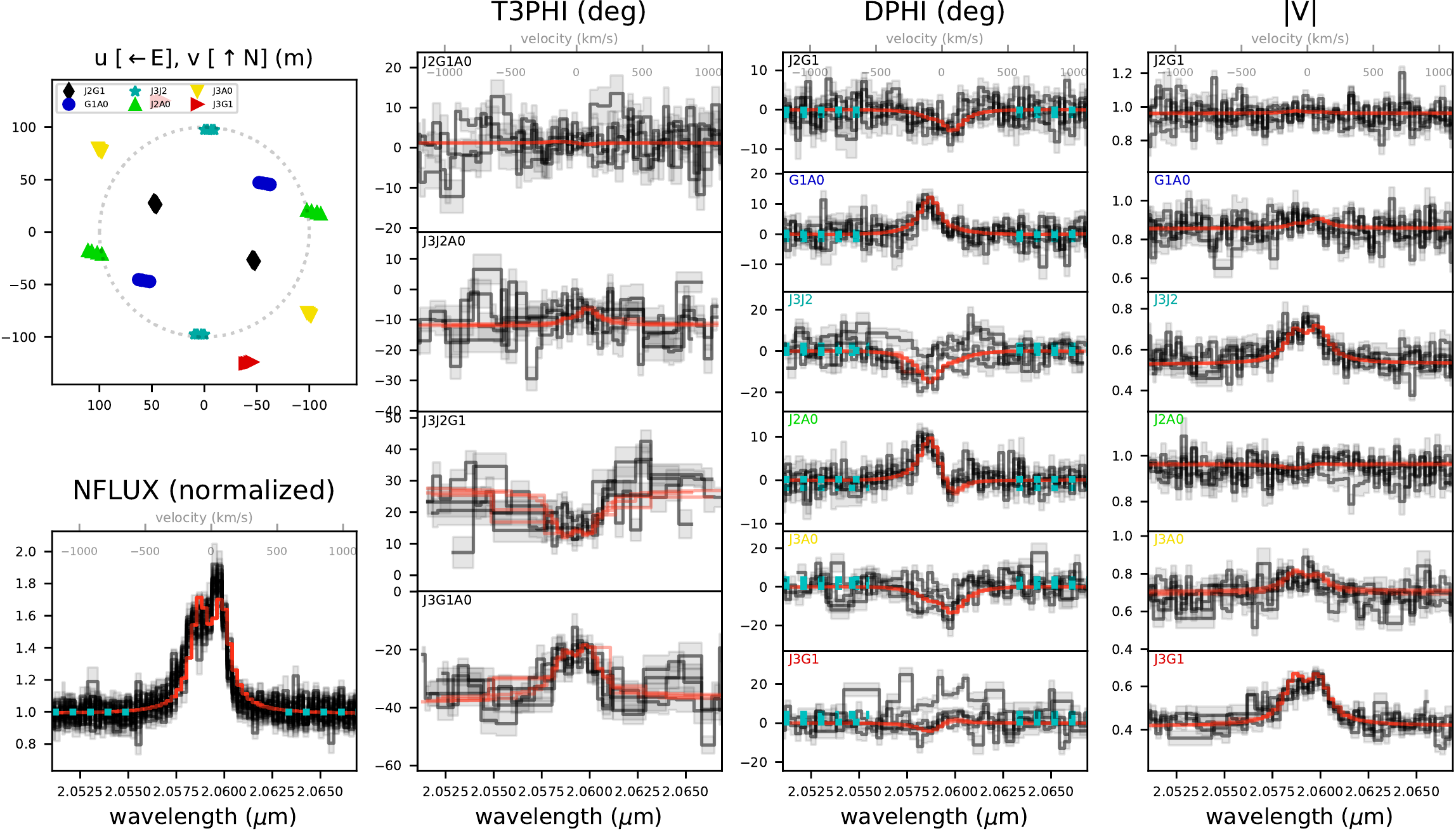}
\includegraphics[width=\textwidth]{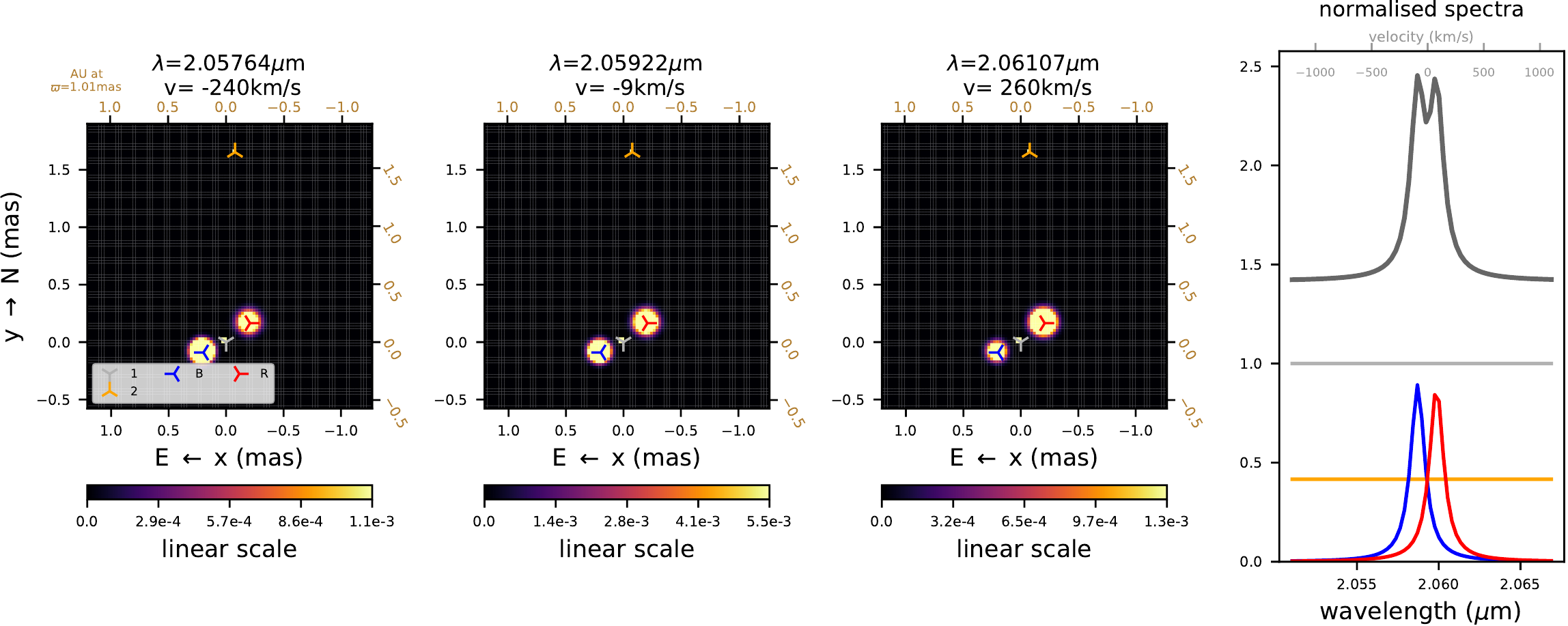}
\caption{2017-03-14, He\,\textsc{i}}
\end{figure}

\begin{figure}
\centering
\includegraphics[width=\textwidth]{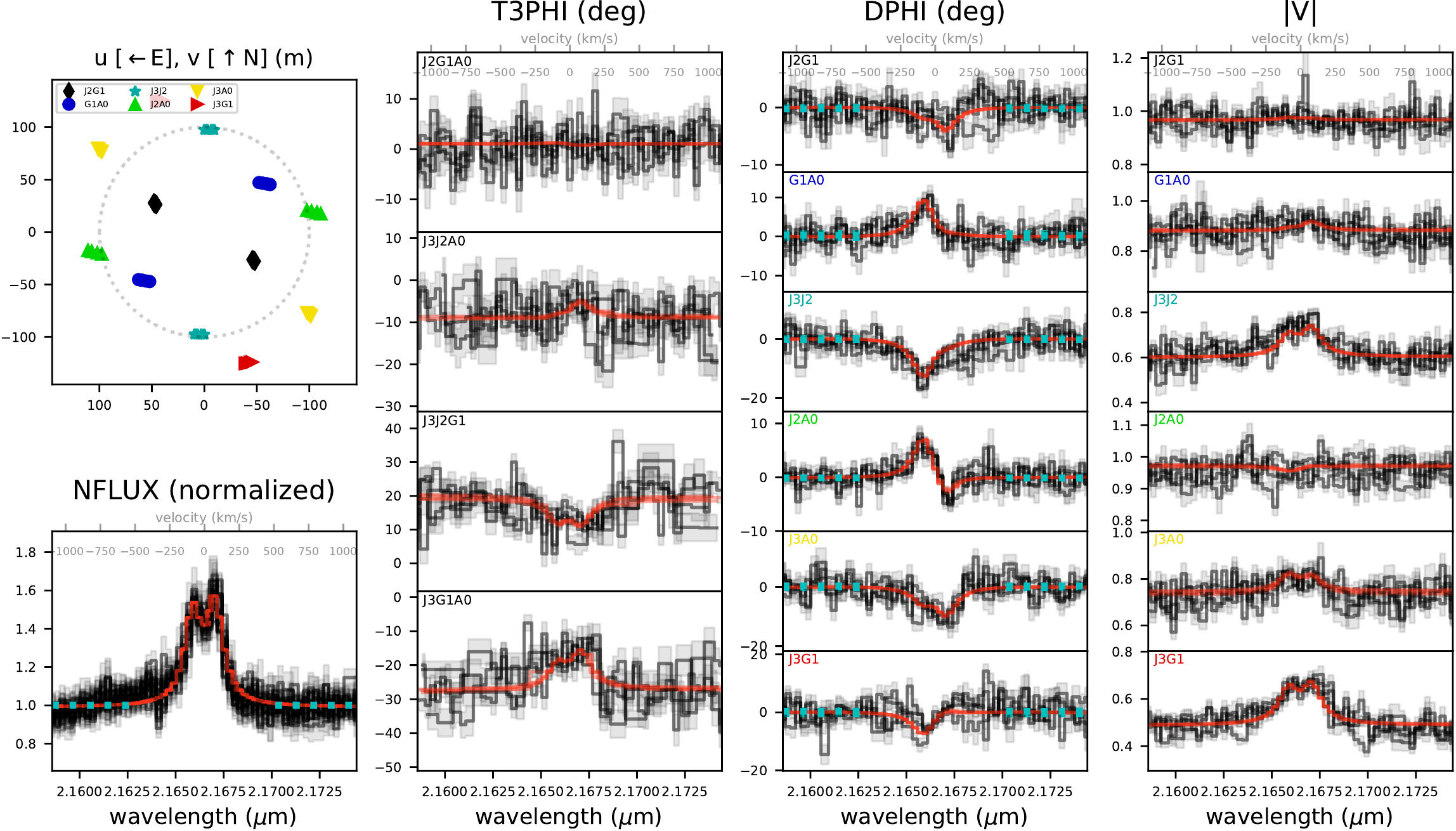}
\includegraphics[width=\textwidth]{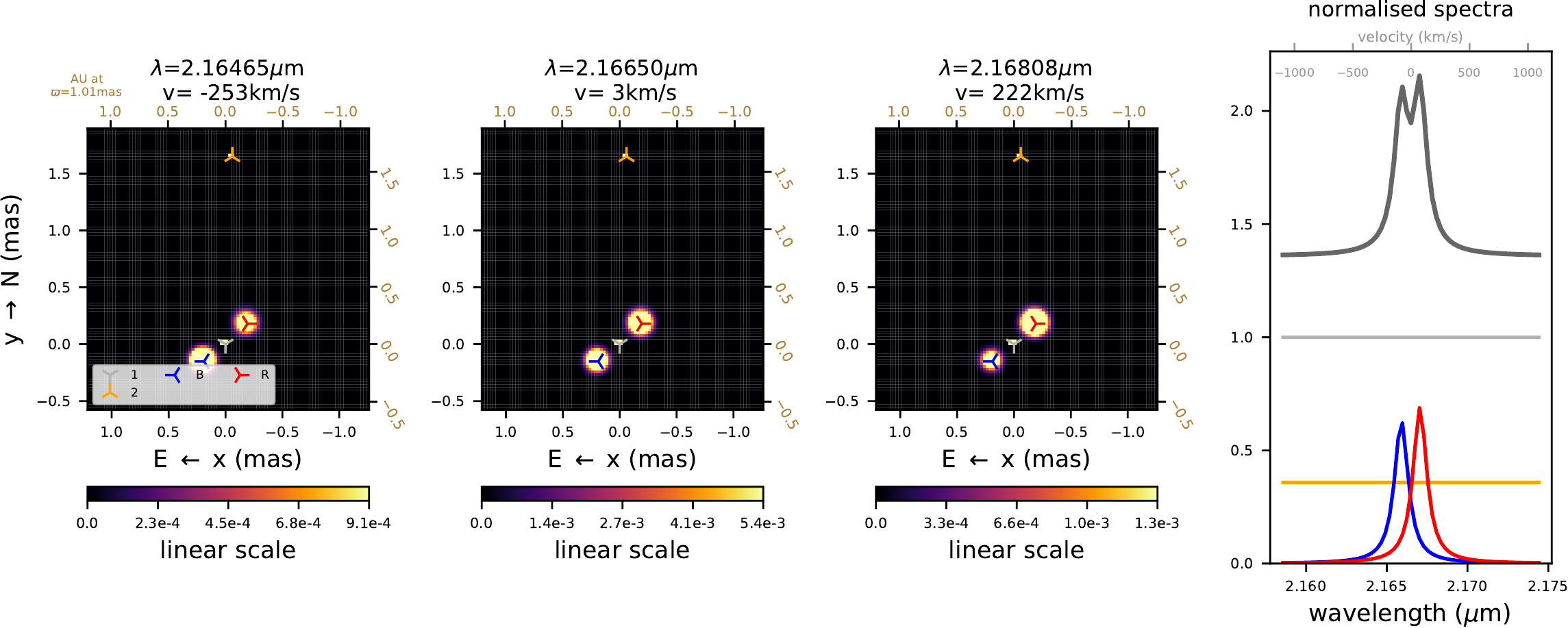}
\caption{2017-03-14, Br\,$\gamma$}
\end{figure}

\newpage

\begin{figure}[h]
\centering
\includegraphics[width=\textwidth]{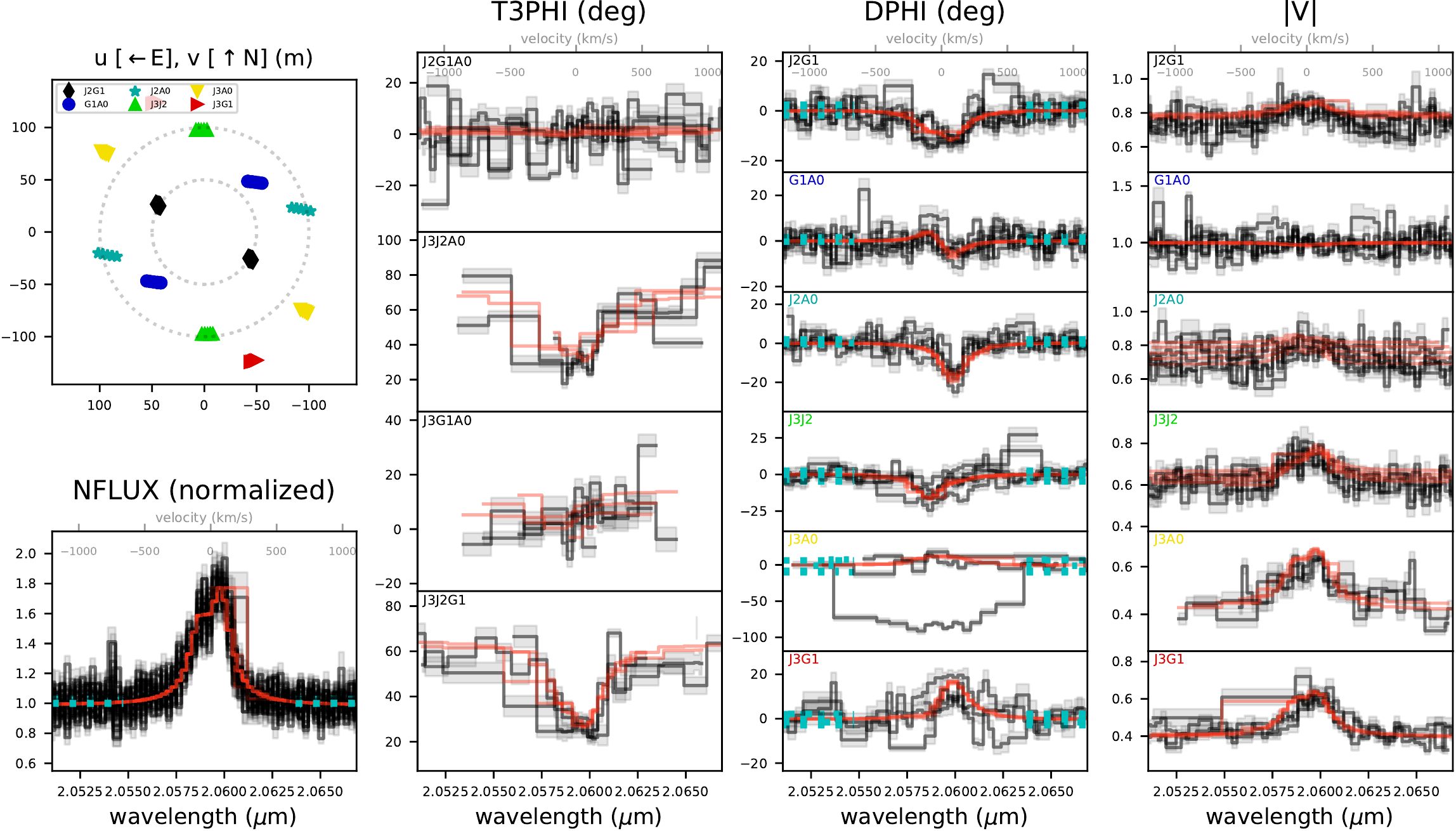}
\includegraphics[width=\textwidth]{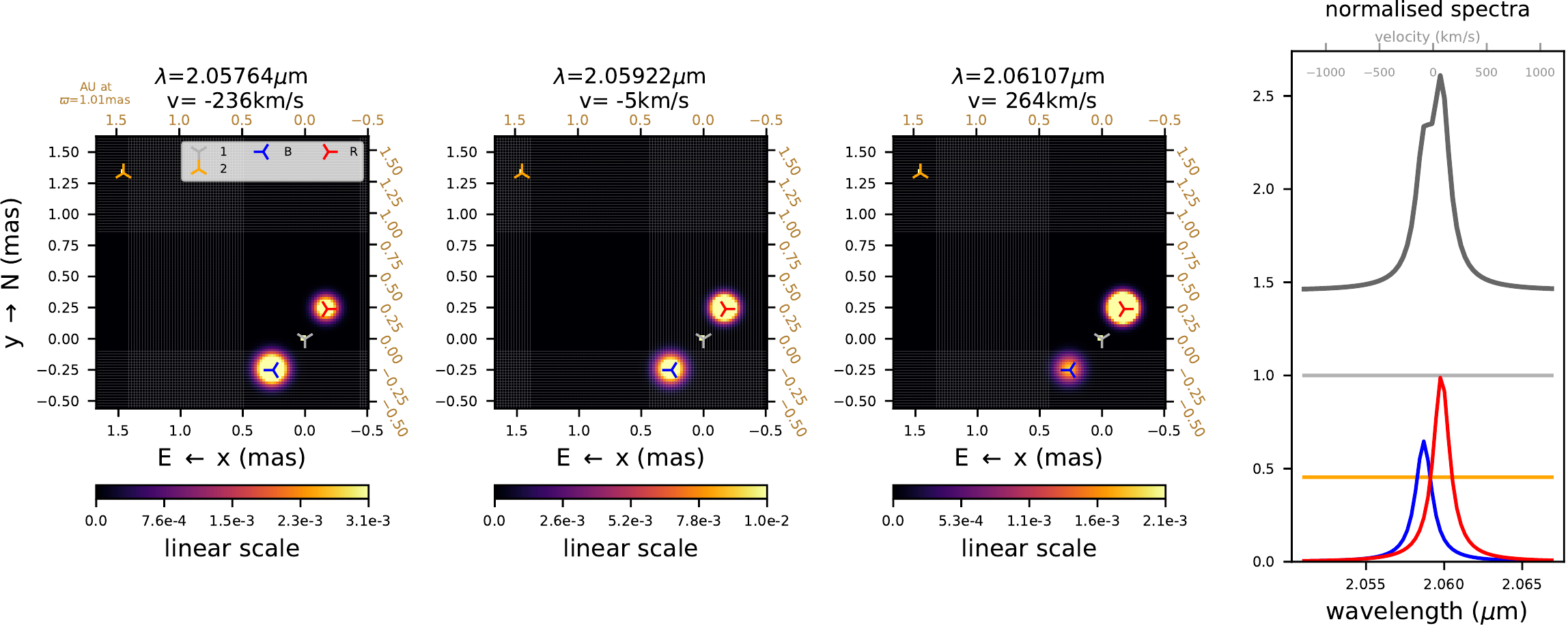}
\caption{2017-04-27, He\,\textsc{i}}
\end{figure}

\begin{figure}[h]
\centering
\includegraphics[width=\textwidth]{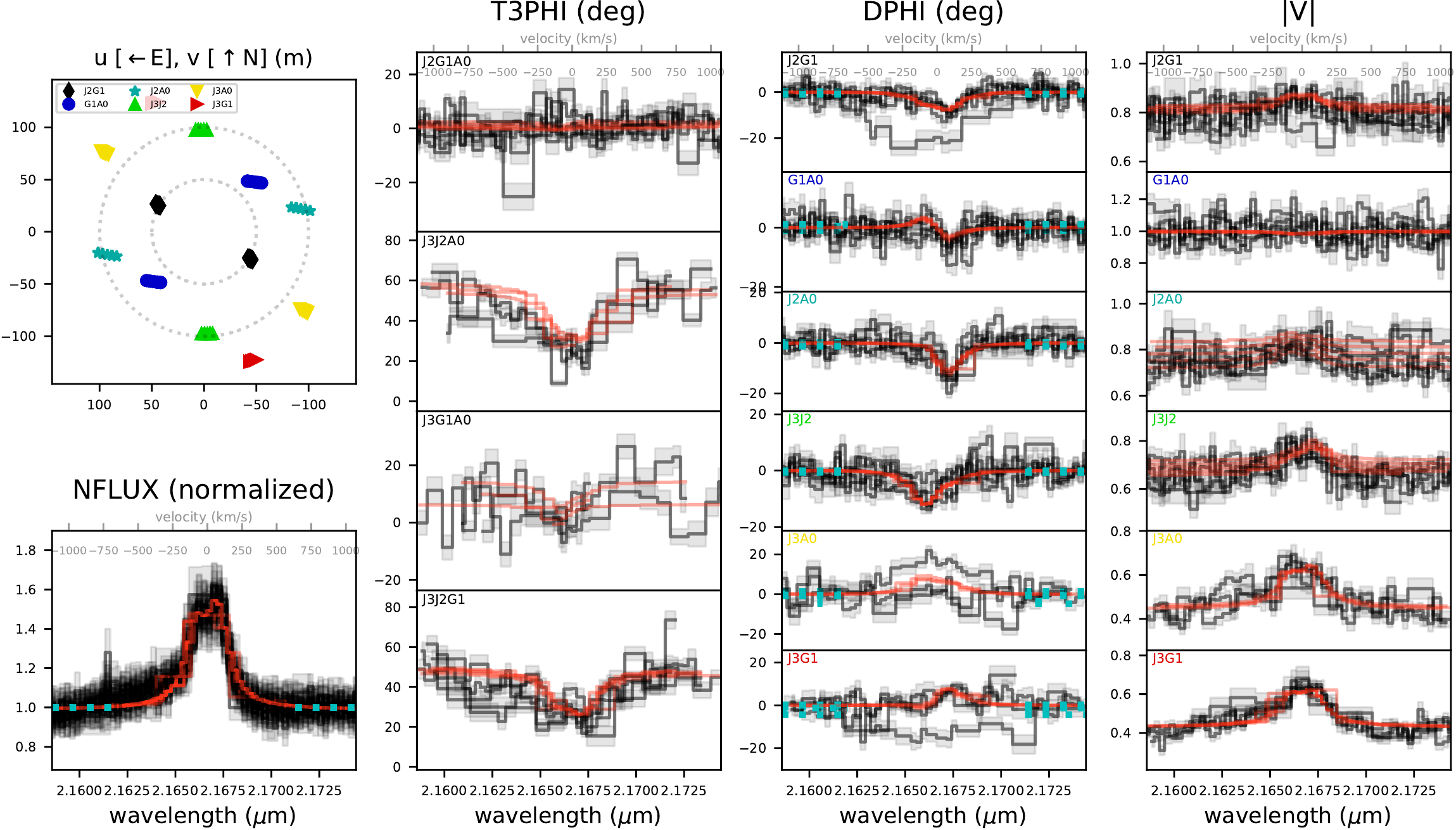}         \includegraphics[width=\textwidth]{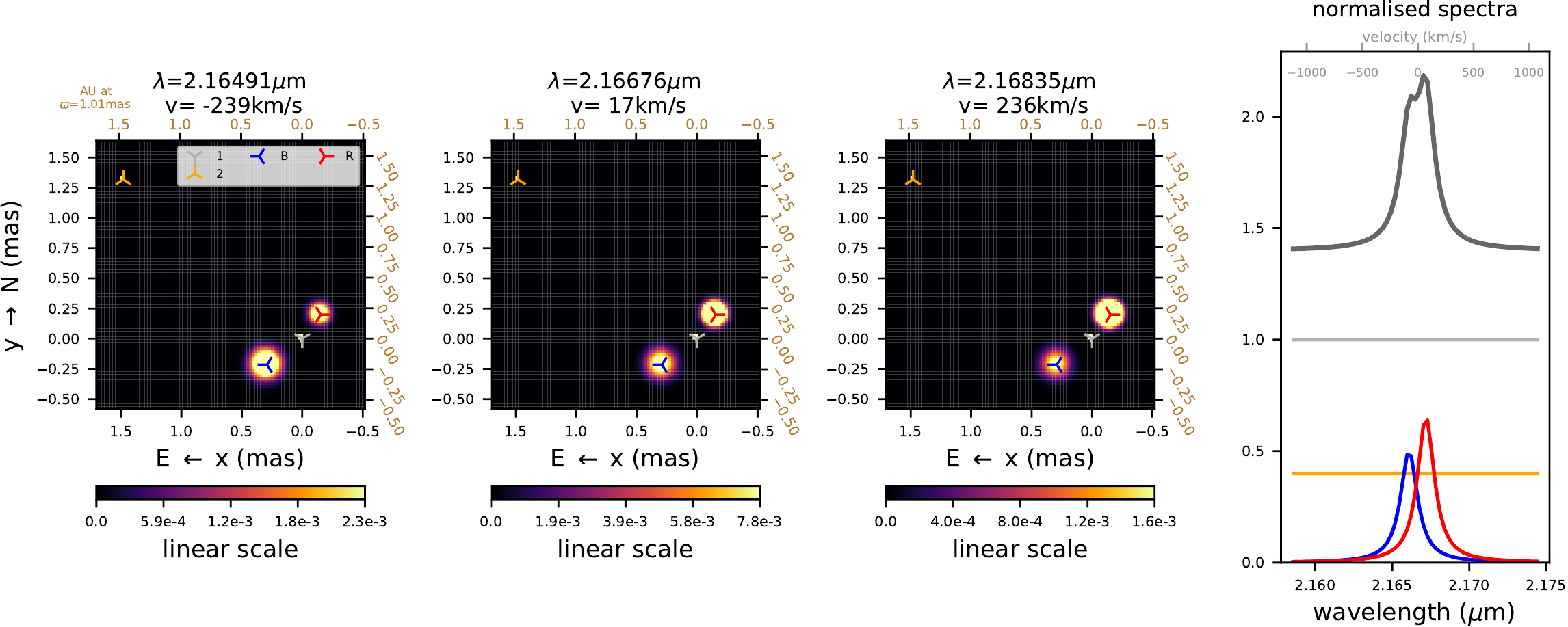}
\caption{2017-04-27, Br\,$\gamma$}
\end{figure}

\begin{figure}[h]
\centering
\includegraphics[width=\textwidth]{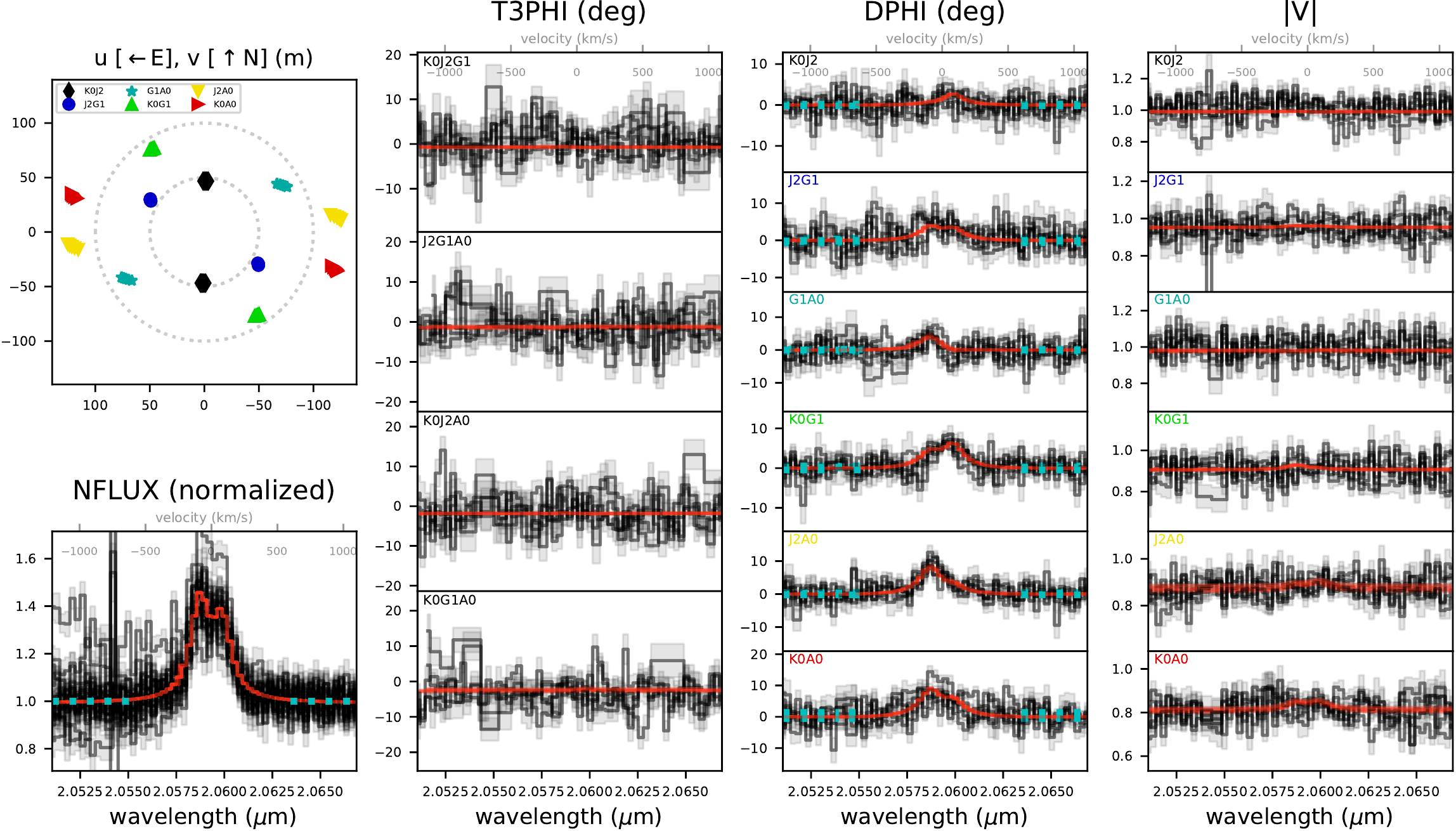}
\includegraphics[width=\textwidth]{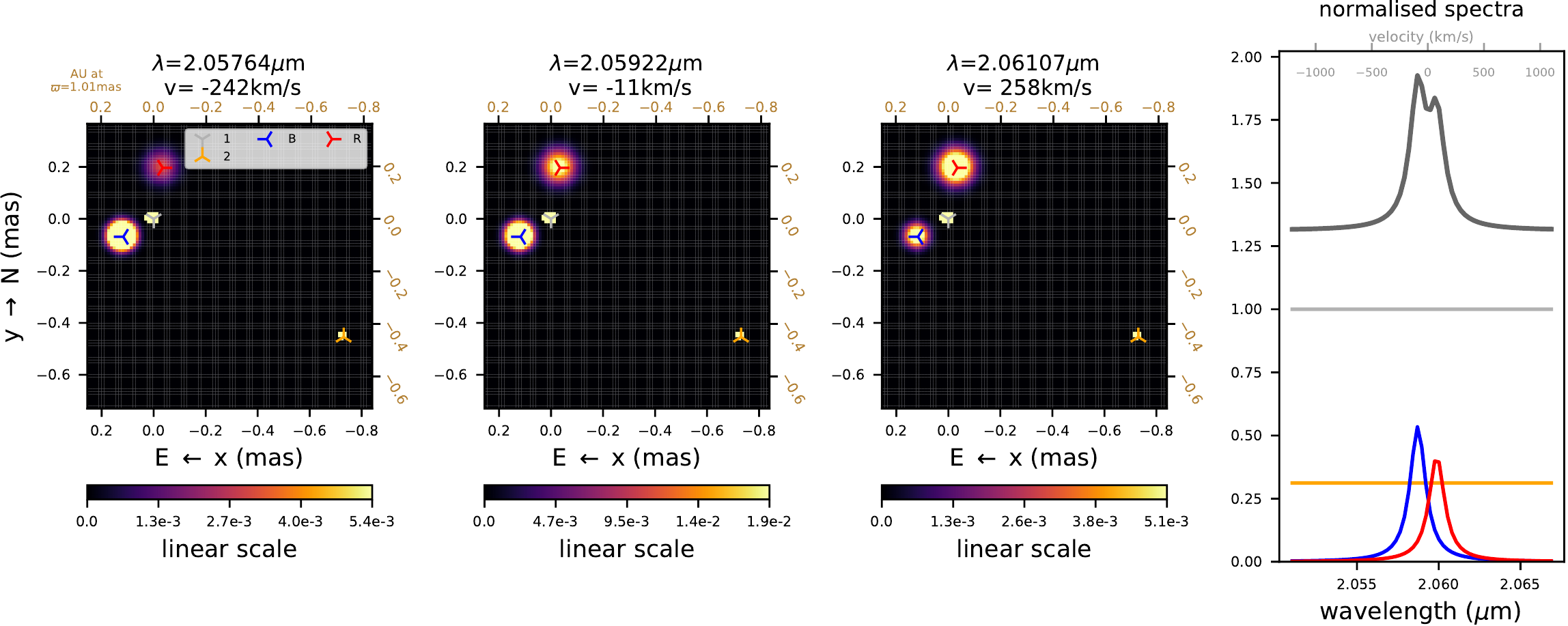}
\caption{2018-01-11, He\,\textsc{i}}
\end{figure}

\begin{figure}[h]
\centering
\includegraphics[width=\textwidth]{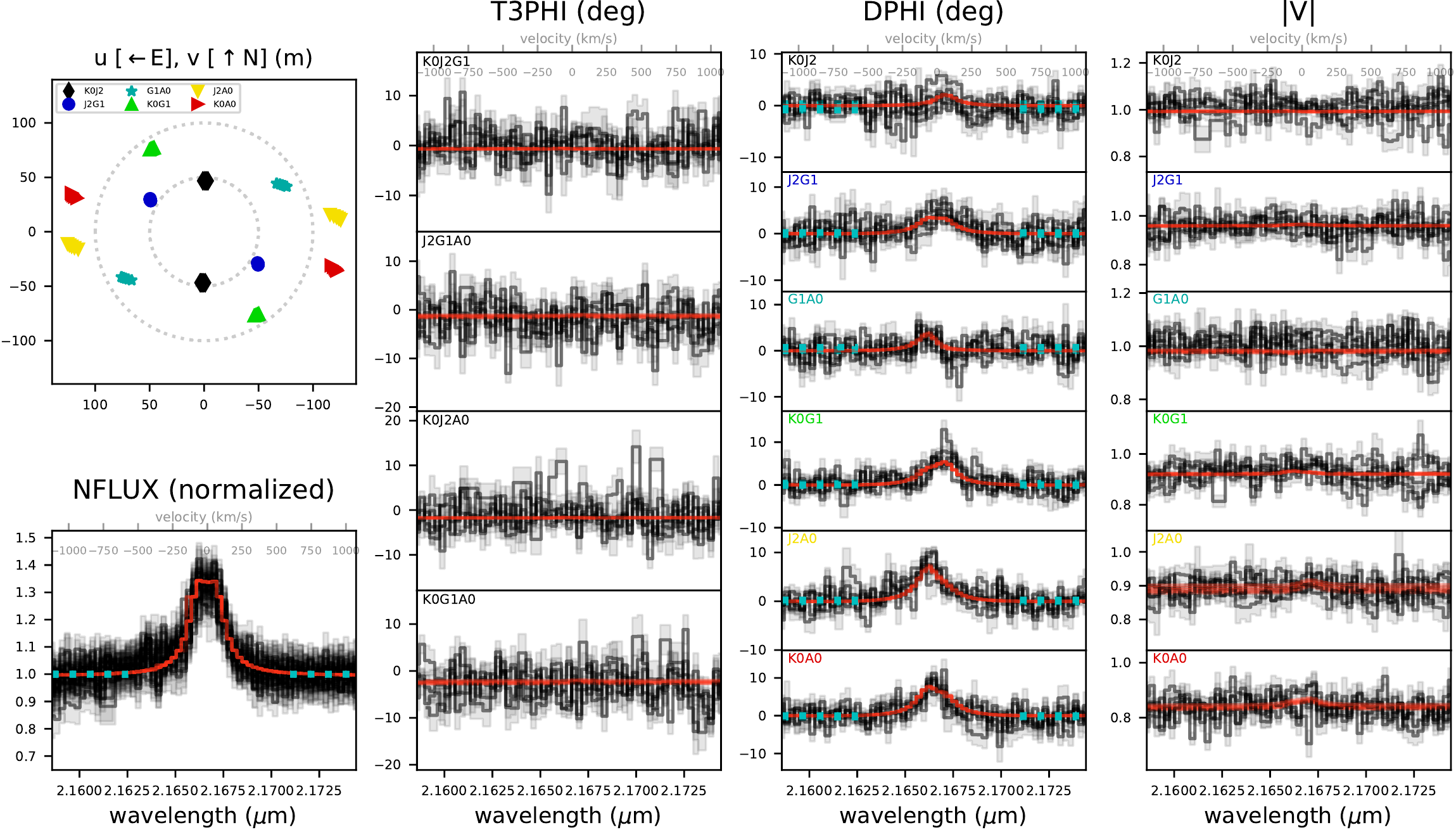}
\includegraphics[width=\textwidth]{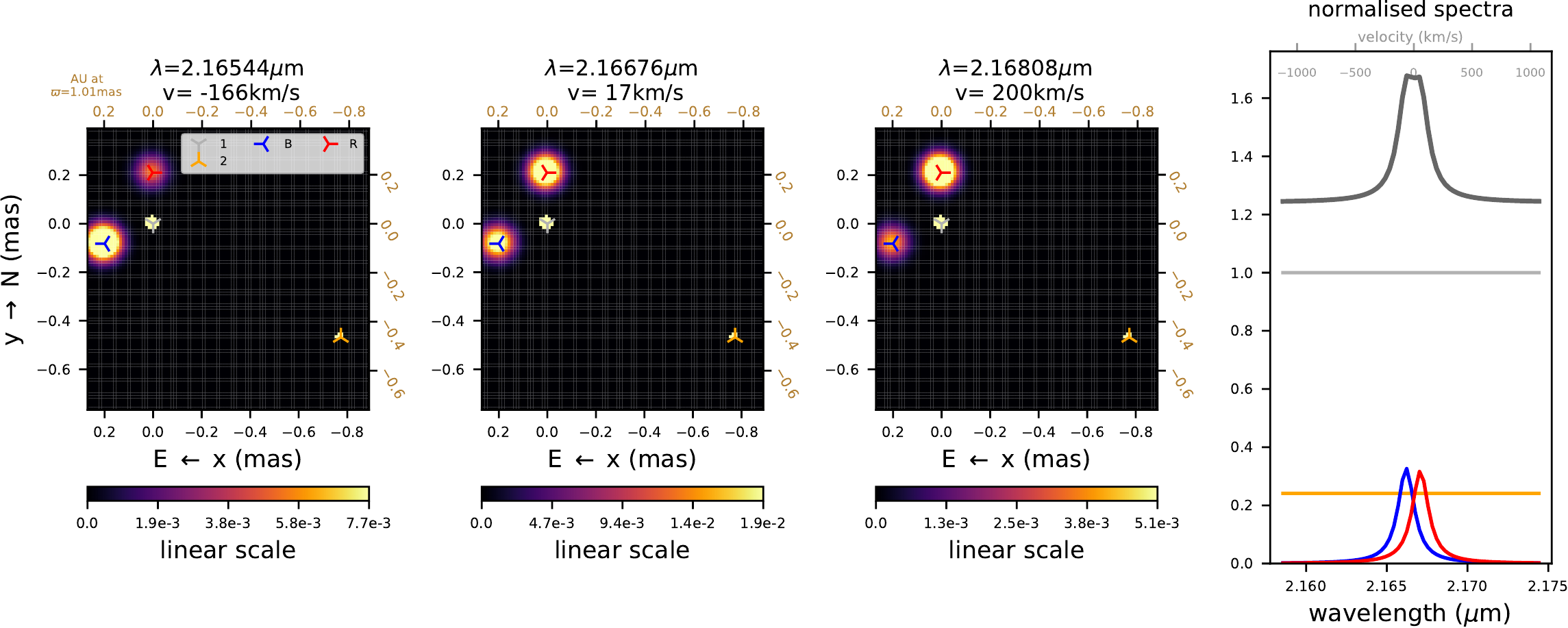}
\caption{2018-01-11, Br\,$\gamma$}
\end{figure}

\newpage
\begin{figure}[h]
\centering
\includegraphics[width=\textwidth]{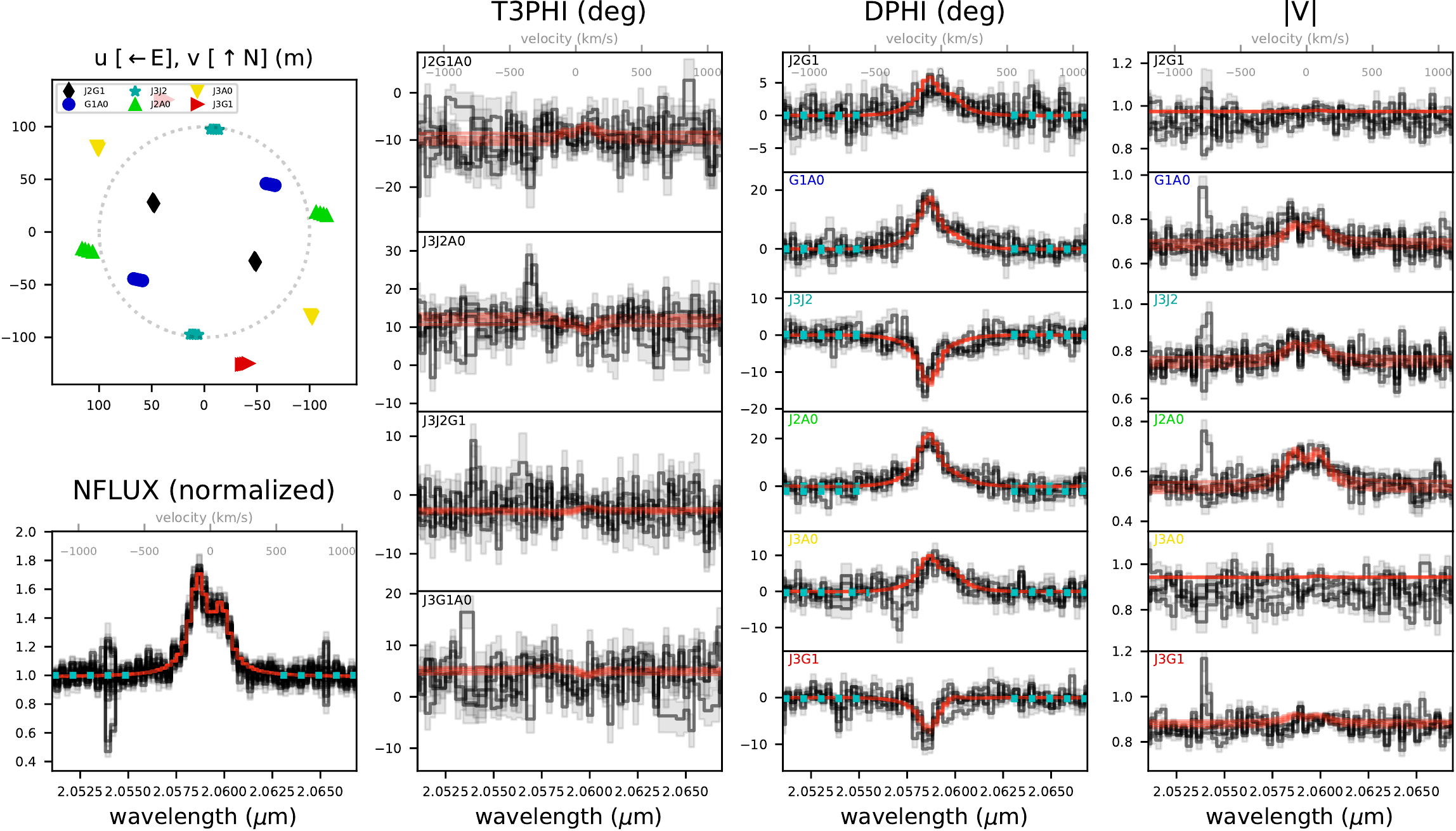}
\includegraphics[width=\textwidth]{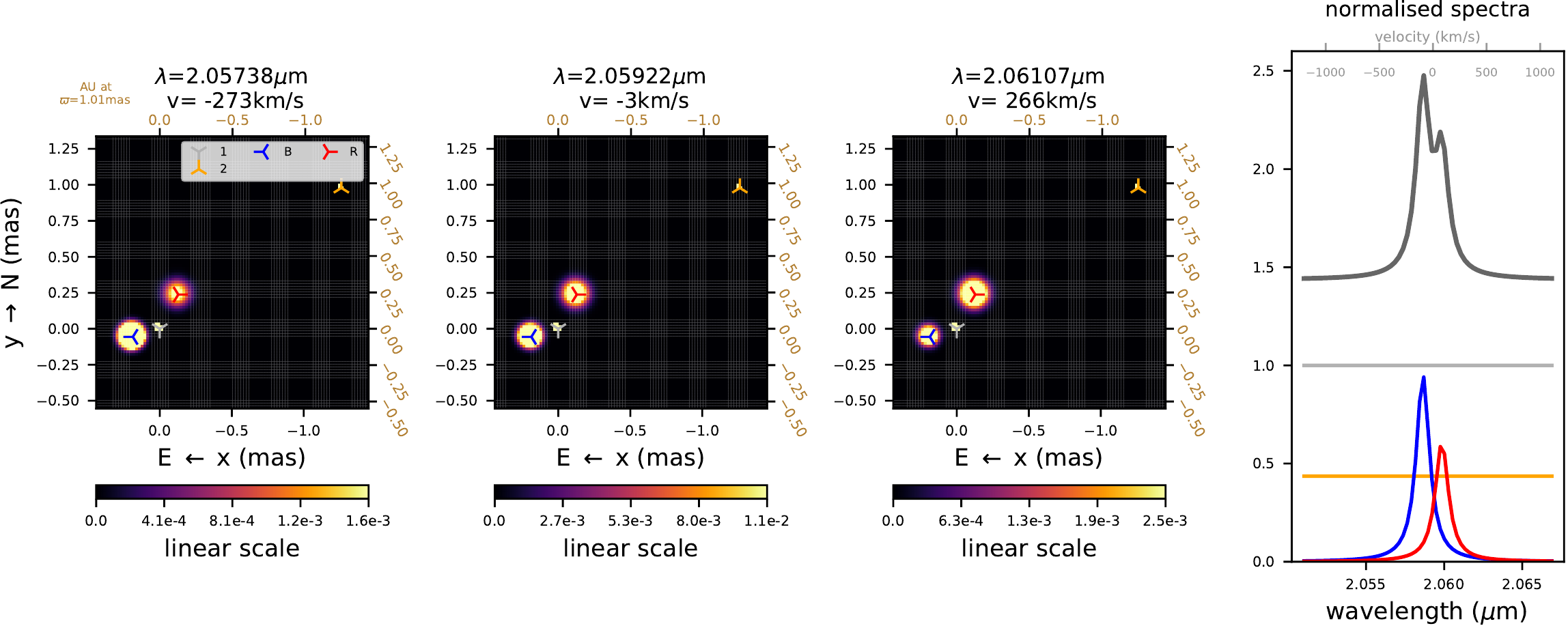}
\caption{2018-02-06, He\,\textsc{i}}
\end{figure}

\begin{figure}[h]
\centering
\includegraphics[width=\textwidth]{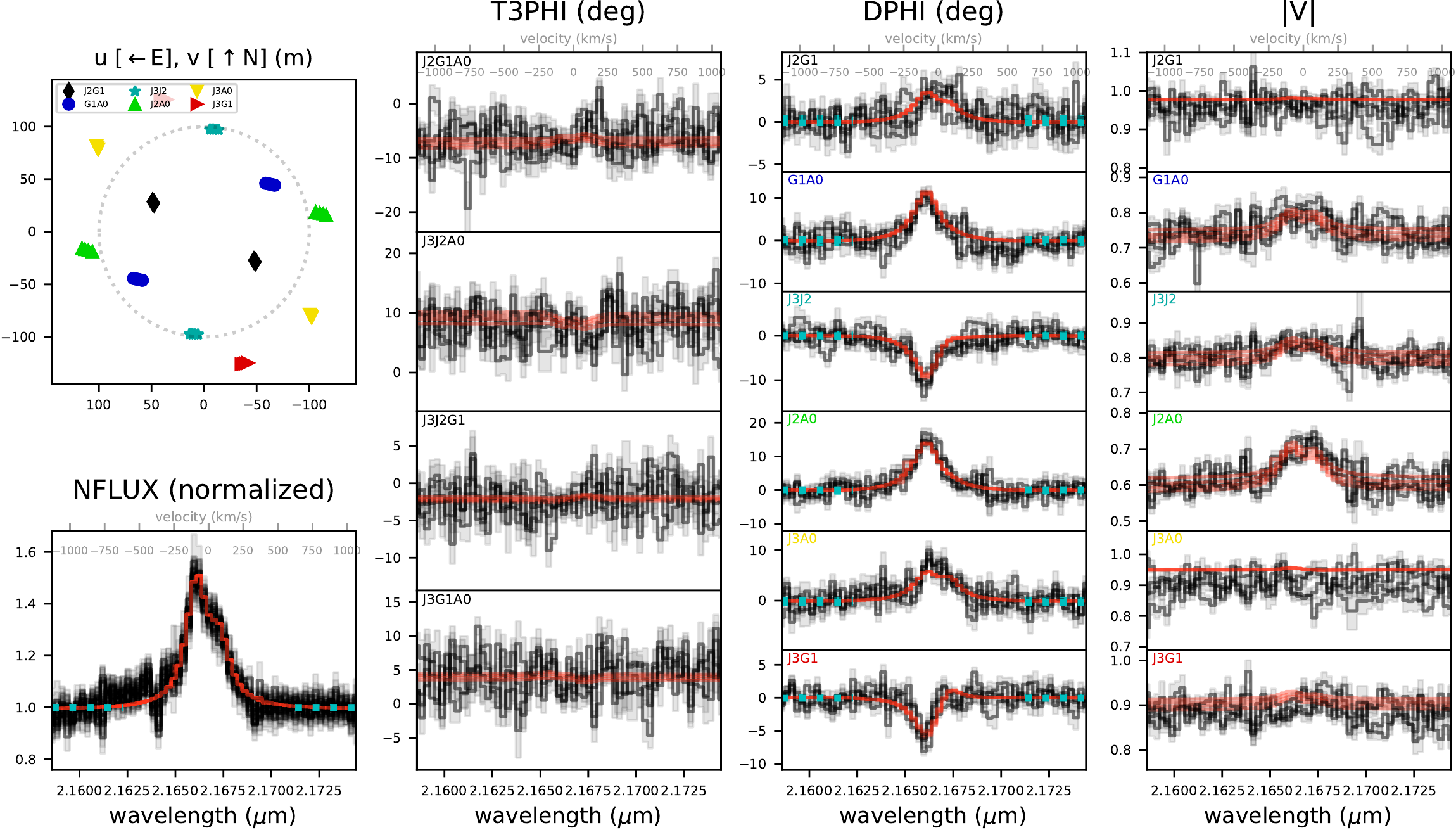}
\includegraphics[width=\textwidth]{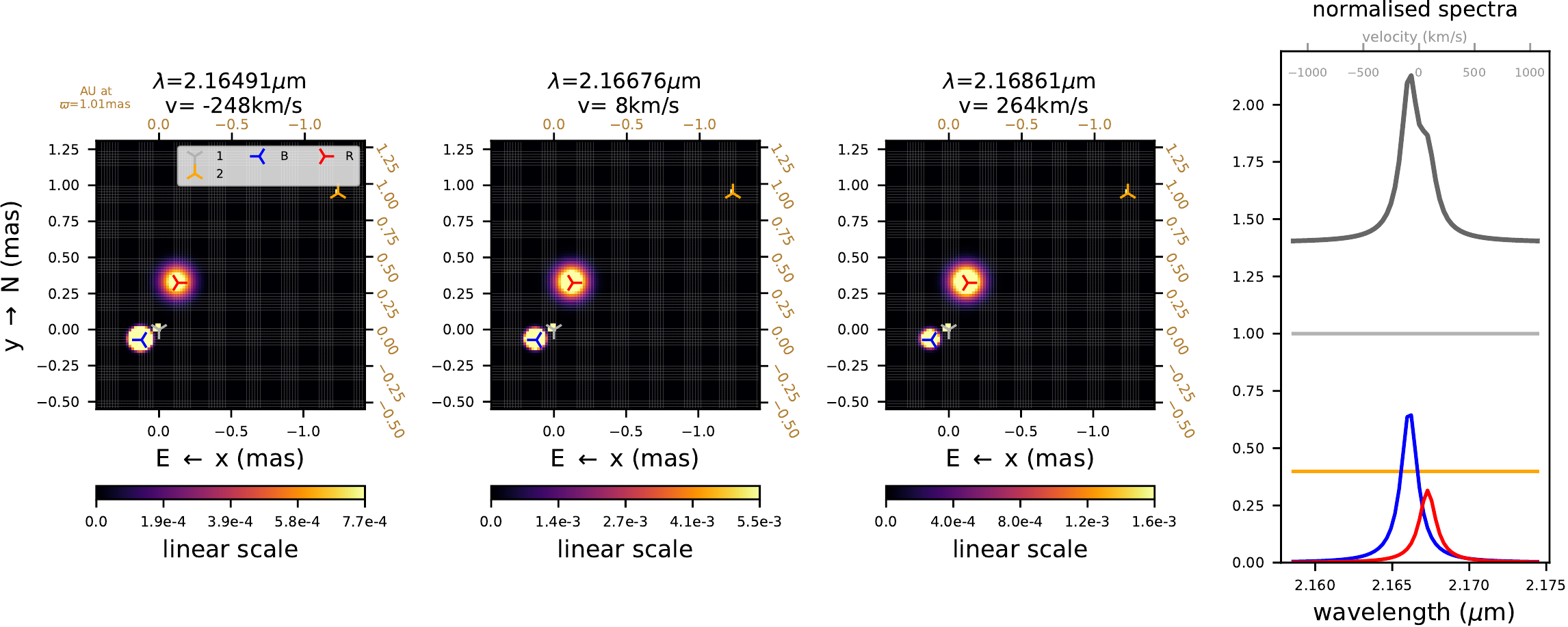}
\caption{2018-02-06, Br\,$\gamma$}
\end{figure}

\end{document}